\documentclass[a4paper,fleqn]{cas-sc}

\usepackage[numbers,sort&compress]{natbib}

\usepackage{fixltx2e}
\usepackage{hyperref}

\usepackage{kotex}

\def\tsc#1{\csdef{#1}{\textsc{\lowercase{#1}}\xspace}}
\tsc{WGM}
\tsc{QE}
\tsc{EP}
\tsc{PMS}
\tsc{BEC}
\tsc{DE}

\begin{document}
\let\WriteBookmarks\relax
\def\floatpagepagefraction{1}
\def\textpagefraction{.001}
\shorttitle{Comprehensive analysis of optimized near-field tandem thermophotovoltaic system}
\shortauthors{J. Song et~al.}

\title [mode = title]{Comprehensive analysis of optimized near-field tandem thermophotovoltaic system}                      



\author[1,2]{Jaeman Song}
\credit{Conceptualization, Methodology, Validation, Data curation, Formal analysis, Writing - original draft}

\author[1,2]{Minwoo Choi}
\credit{Methodology, Validation, Writing - review \& editing}

\author[3]{Mikyung Lim}
\credit{Methodology, Validation, Writing - review \& editing}

\author[1,2]{Jungchul Lee}
\credit{Supervision, Writing - review \& editing}

\author[1,2]{Bong Jae Lee}
\cormark[1]
\fnmark[1]
\credit{Conceptualization, Supervision, Writing - review \& editing, Funding acquisition}

\address[1]{Department of Mechanical Engineering, Korea Advanced Institute of Science and Technology, Daejeon 34141, South Korea}
\address[2]{Center for Extreme Thermal Physics and Manufacturing, Korea Advanced Institute of Science and Technology, Daejeon 34141, South Korea}
\address[3]{Nano-Convergence Mechanical Systems Research Division, Korea Institute of Machinery and Materials, Daejeon 34103, South Korea}

\cortext[cor1]{Corresponding author}

\fntext[fn1]{E-mail address: bongjae.lee@kaist.ac.kr}

\begin{abstract}
It is well known that performance of a thermophotovoltaic (TPV) device can be enhanced if the vacuum gap between the thermal emitter and the TPV cell becomes nanoscale due to the photon tunneling of evanescent waves. Having multiple bandgaps, multi-junction TPV cells have received attention as an alternative way to improve its performance by selectively absorbing the spectral radiation in each subcell. In this work, we comprehensively analyze the optimized near-field tandem TPV system consisting of the thin-ITO-covered tungsten emitter (at 1500 K) and GaInAsSb/InAs monolithic interconnected tandem TPV cell (at 300 K). We develop a simulation model by coupling the near-field radiation solved by fluctuational electrodynamics and the diffusion-recombination-based charge transport equations. The optimal configuration of the near-field tandem TPV system obtained by the genetic algorithm achieves the electrical power output of 8.41 W/cm$^2$ and the conversion efficiency of 35.6\% at the vacuum gap of 100 nm. We show that two resonance modes (i.e.,  surface plasmon polaritons supported by the ITO-vacuum interface and the confined waveguide mode in the tandem TPV cell) greatly contribute to the enhanced performance of the optimized system. We also show that the near-field tandem TPV system is superior to the single-cell-based near-field TPV system in both power output and conversion efficiency through loss analysis. Interestingly, the optimization performed with the objective function of the conversion efficiency leads to the current matching condition for the tandem TPV system regardless of the vacuum gap distances.
\end{abstract}


\maketitle

\section{Introduction}
A thermophotovoltaic (TPV) system consists of a high-temperature emitter and a TPV cell, and directly converts the radiative energy into the electrical energy through a photovoltaic effect \cite{park2013fundamentals, datas2017thermophotovoltaic, tervo2018near, datas2021thermophotovoltaic}. One of the advantages of TPV energy conversion is its versatility of the thermal source, leading that it can be utilized wherever the emitter could be heated to high temperatures. For example, when the TPV system is applied in industries, the waste heat can be recovered to the electrical energy  \cite{park2013fundamentals, zhao2017high, tervo2018near, datas2021thermophotovoltaic}. Moreover, the TPV system can also be applied to a full-spectrum solar energy harvesting system named a solar TPV (STPV) system \cite{nam2014solar, shimizu2015high, ni2019theoretical}. Compared to the conventional solar cell, the STPV system has a potential to surpass the Shockley-Queisser (SQ) limit of a single-junction solar cell because it can substantially reduce the angular mismatch loss. In addition, the TPV system retains advantages of solid-state devices, such as compactness and simplicity in configuration, noise- and vibration-free operation, and potential for miniaturization \cite{park2013fundamentals, datas2017thermophotovoltaic, tervo2018near, datas2021thermophotovoltaic}.

It is well known that the performance of a TPV system can be enhanced when a gap between the emitter and the TPV cell is smaller than a thermal characteristic wavelength determined by Wien’s displacement law \cite{park2008performance, francoeur2011thermal, park2013fundamentals, bright2014performance, lim2015graphene, chang2015tungsten, tong2015thin, jin2016hyperbolic, zhao2017high, yang2017performance, vongsoasup2017performance, st2017hot, lim2018optimization, tervo2018near, vaillon2019micron, sabbaghi2019near, song2019analysis, liao2019graphene, papadakis2020broadening, sabbaghi2020near, datas2021thermophotovoltaic, inoue2021near}. In this near-field TPV (NFTPV) system, thermal radiation exceeding the blackbody limitation can be transferred to the TPV cell due to the contribution of evanescent waves. To further enhance the electrical power generation and the conversion efficiency of the NFTPV system, several works have been focused on optical tuning approach; that is, tailoring the spectrum of thermal radiation by modifying the surface of the emitter or the TPV cell using nanostructures \cite{lim2015graphene, chang2015tungsten, tong2015thin, jin2016hyperbolic, zhao2017high, yang2017performance, vongsoasup2017performance, st2017hot, lim2018optimization, sabbaghi2019near, papadakis2020broadening}. For example, by introducing hyperbolic metamaterials (HMMs) as an emitter (e.g., nanowire \cite{chang2015tungsten}, multilayer \cite{jin2016hyperbolic, lim2018optimization}, and grating \cite{vongsoasup2017performance}), it was demonstrated that hyperbolic modes supported by HMMs can improve the performance of the NFTPV system. Moreover, multilayers \cite{yang2017performance, lim2018optimization, papadakis2020broadening} and thin films \cite{lim2015graphene, tong2015thin, zhao2017high, st2017hot} can be introduced to the emitter or the TPV cell cover, exciting the surface plasmon polaritons (SPPs) supported inside the layered structure. It was also reported that magnetic polariton (MP) excited by the grating emitter can also tailor the emission spectrum of the NFTPV system \cite{sabbaghi2019near}. 

Alternatively, one can change the configuration of the TPV cell to further improve its performance \cite{bright2014performance, tong2015thin, st2017hot, lim2018optimization, sabbaghi2019near, song2019analysis, liao2019graphene, papadakis2020broadening, sabbaghi2020near, inoue2021near}. When an infrared reflector is applied to the backside of the TPV cell, the conversion efficiency of the NFTPV system can be significantly increased by reducing the absorption of the sub-bandgap radiation in the TPV cell \cite{bright2014performance, tong2015thin, lim2018optimization, sabbaghi2019near, papadakis2020broadening}. Recently, it was also found that the non-contact reflector can promote efficient photon recycling \cite{inoue2021near}. In addition, several groups proposed the thin-film TPV cell confined by a backside reflector and demonstrated the performance enhancement of the NFTPV system through waveguide modes \cite{tong2015thin}, bulk polariton \cite{lim2018optimization}, and MP \cite{sabbaghi2019near}. As an alternative to a \textit{p-n} junction TPV cell, a metal-semiconductor Schottky junction TPV cell was also employed in the NFTPV system with an advantage of a wider spectral response range compared to the \textit{p-n} TPV cell \cite{st2017hot, song2019analysis, liao2019graphene}. 

Because a multi-junction cell can surpass the SQ limit of a single-junction cell, multi-junction cells have been intensively investigated in various photovoltaic fields from theoretical design to real fabrication \cite{geisz200840, france2015design, dimroth2014wafer, france2016metamorphic, cariou2016monolithic, hossain2016novel, lou2017enhanced, datas2017monolithic, philipps2018high, sabbaghi2020near, kret2021investigation}. Subcells constituting the multi-junction cell are arranged in the order of decreasing the bandgap from the top where the thermal radiation is incident, and each subcell is monolithic interconnected by tunnel diodes. Through this configuration, each subcell can absorb the spectral radiation close to its bandgap leading to the substantially reduced thermalization loss.  \citet{sabbaghi2020near} firstly reported the near-field tandem TPV system composed of thin GaSb and InAs subcells. In their tandem cell, each subcell operates individually, meaning that subcells are mechanically stacked. Because of the high cost of the mechanically stacked multi-junction cell, recent works have focused on developing the monolithic multi-junction cell \cite{philipps2018high}. However, little has been done on further improving the performance of near-field tandem TPV system by employing the aforementioned optical tuning approaches. 

In this work, we throughly analyze the performance of the near-field tandem TPV system consisting of a monolithic interconnected InAs and GaInAsSb subcells and a thin-ITO-covered emitter. A simulation model coupling the near-field thermal radiation based on fluctuating electrodynamics and the charge transport equations is developed. Then, we optimize the configuration of the proposed NFTPV system using a genetic algorithm to maximize the conversion efficiency at a given vacuum gap distance. The performance of the single-cell-based and the tandem-cell-based NFTPV systems will be comprehensively compared through the loss analysis. In addition, by optimizing the system structure at different vacuum gap distances using the same design variables, we confirm whether the proposed NFTPV system could find a high-performance configuration that satisfies the current matching condition regardless of the vacuum gap distance.

\begin{figure}[!b]
\centering
\includegraphics[width=0.7\textwidth]{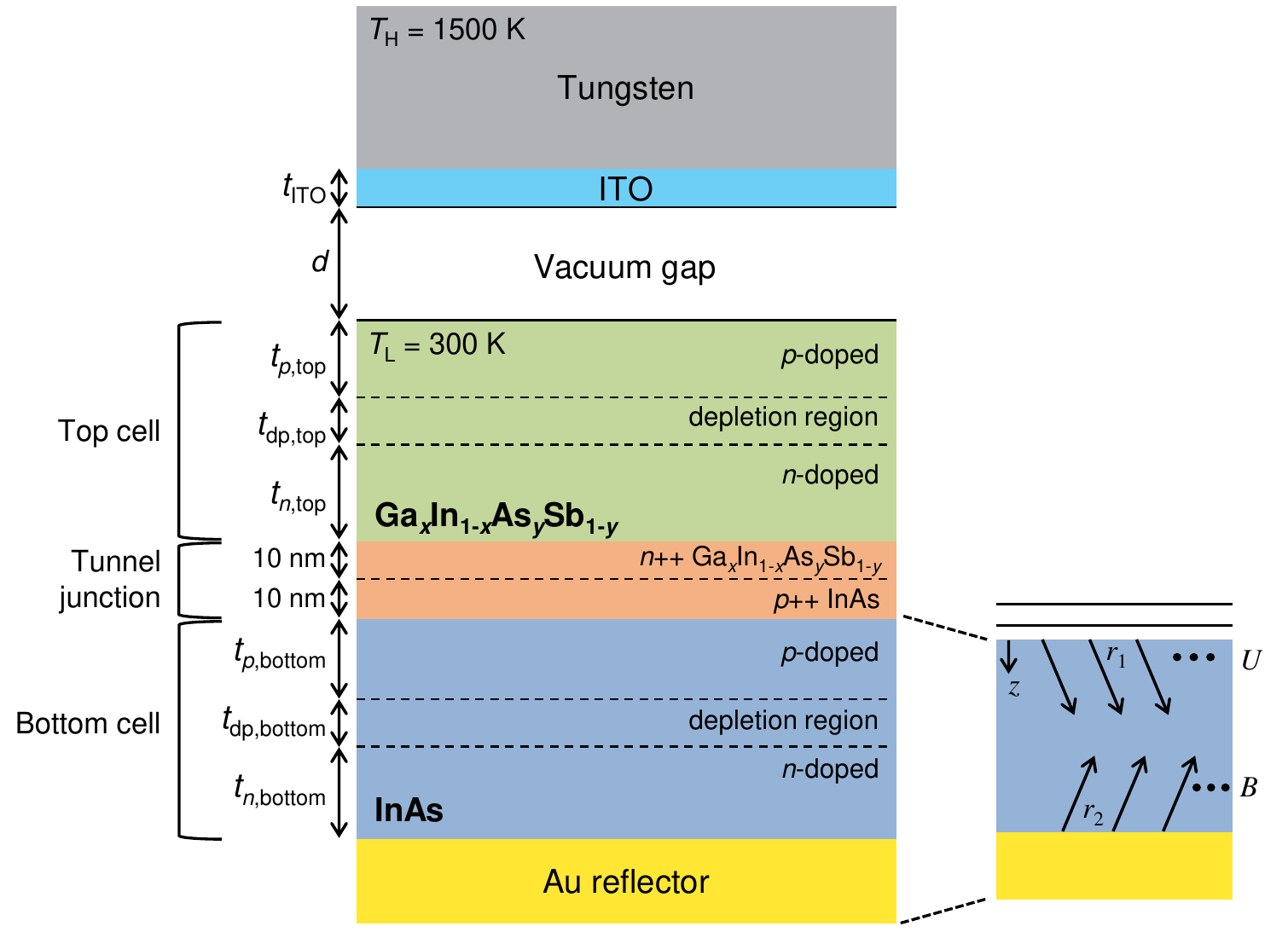}
\caption{\label{fig1} 
Schematic illustration of the proposed near-field tandem thermophotovoltaic system. }
\end{figure}
%

\section{Theoretical modeling}

Figure \ref{fig1} illustrates a schematic of the NFTPV system consisting of a thin-ITO-covered tungsten emitter and a tandem TPV cell. Temperatures of the emitter and the tandem TPV cell are set to be 1500 K and 300 K, respectively. The vacuum gap between them is fixed to 100 nm except for the cases discussed in the last two paragraphs in Section 3. Please note that the 100-nm-gap between the emitter and the TPV cell is an achievable gap considering the experimentally validated state-of-the-art NFTPV systems \cite{fiorino2018nanogap, inoue2019one, lucchesi2021near, bhatt2020integrated}.

ITO has been widely adapted to NFTPV works because its plasma frequency can be tuned by regulating the oxygen concentration during the fabrication process \cite{gregory2002high, jeong2008effect, lee2014nanoscale, zhao2017high, st2017hot, papadakis2020broadening}. Since the resonance frequency of SPPs supported at the ITO-vacuum interface depends on the plasma frequency of ITO, the emission spectrum can be readily controlled by ITO. Moreover, both ITO and tungsten are known to be sustainable at high temperatures \cite{zhao2017high, st2017hot,papadakis2020broadening}.

The tandem cell consists of two $p$-on-$n$ TPV subcells: an InAs bottom cell and a GaInAsSb top cell. Because an InAs cell has a narrow bandgap and can operate at room temperatures \cite{lu2018inas, milovich2020design}, we selected it as a bottom cell. For a wider-bandgap top cell, we chose Ga$_{x}$In$_{1-x}$As$_{y}$Sb$_{1-y}$ quaternary semiconductor of which electrical and optical properties can be tuned according to the $x$ and $y$ composition ratio. We permit the combination of $x$ and $y$ only when the lattice constant of Ga$_{x}$In$_{1-x}$As$_{y}$Sb$_{1-y}$, which is a function of $x$ and $y$, is within 4.5\% mismatch from that of an InAs considering methods to connect of lattice-mismatched materials, such as metamorphic \cite{france2016metamorphic, kret2021investigation}, inverted metamorphic \cite{geisz200840, france2015design}, and direct wafer bonding \cite{dimroth2014wafer, cariou2016monolithic}. The ideal tunnel junction located between two subcells for the monolithic interconnection is assumed to be formed using 10-nm-thick $n$- and $p$-doped semiconductors of each subcell. The total thickness of the tandem TPV cell is limited by the Au electrode placed under the InAs bottom cell. The bottom Au electrode can also act as a backside reflector to reject the sub-bandgap absorption, which is one of the critical factors to deteriorate the performance of the TPV system. Some layers required to fabricate high-quality multi-junction cells, such as the back surface field (BSF), buffer, and anti-oxidation layer, are not considered in this work for simplicity.

\begin{figure}[!b]
\centering
\includegraphics[width=0.5\textwidth]{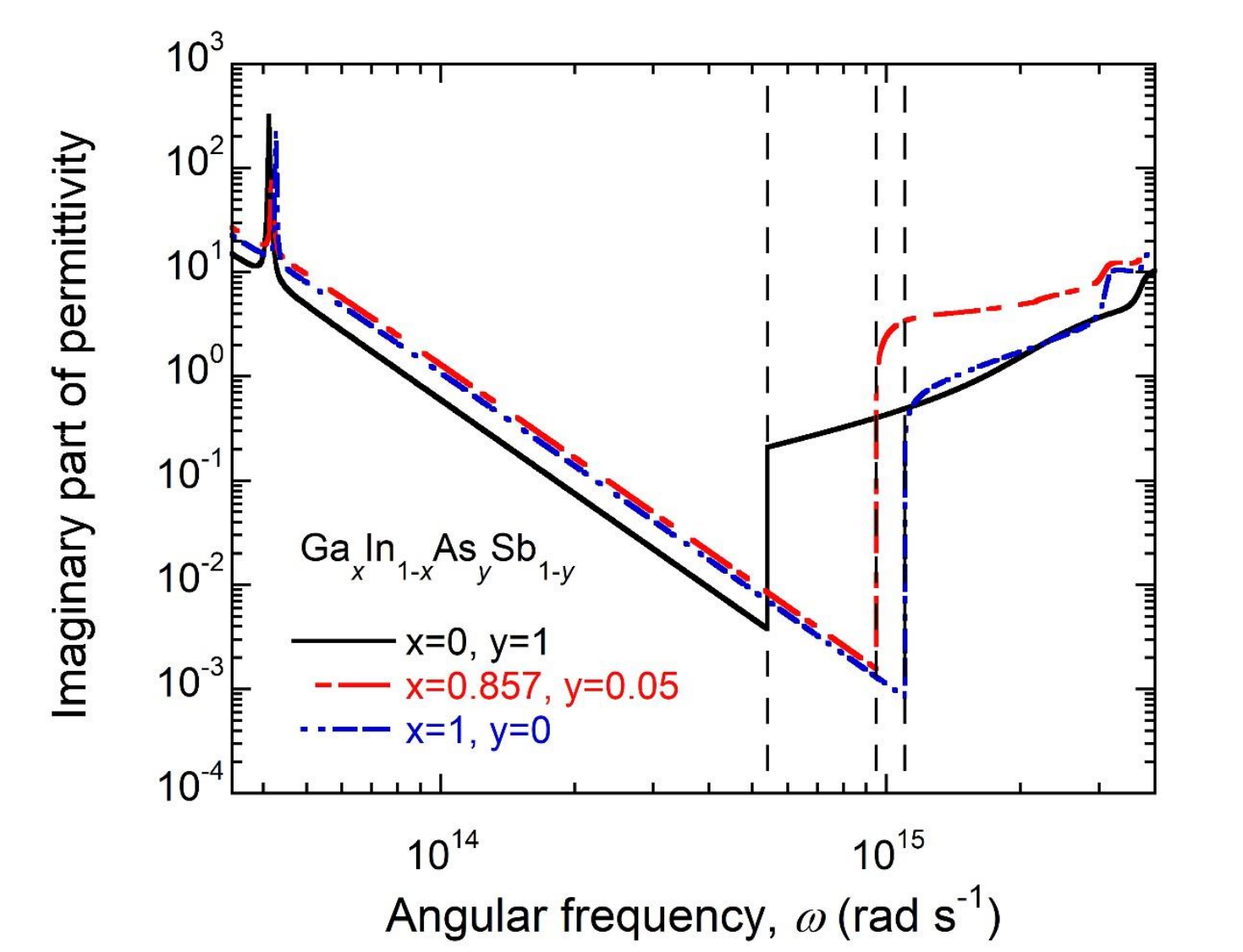}
\caption{\label{fig2} The imaginary part of permittivity of Ga$_x$In$_{1-x}$As$_y$Sb$_{1-y}$ for three $[x,y]$ combinations. Black vertical dashed lines indicate the angular frequency corresponding to the bandgap energy.
}
\end{figure}
%

Dielectric function $\epsilon(x,y,\omega)$ of semiconductors including quaternary compound Ga$_{x}$In$_{1-x}$As$_{y}$Sb$_{1-y}$ above its bandgap is obtained from the semi-empirical model proposed by \cite{adachi1989JAP}, and the Lorentz-Drude (LD) oscillator model \cite{rakic1998optical} is used for the sub-bandgap region. The model parameters and basic properties such as the bandgap frequency and lattice constant depends on the composition ratios $[x,y]$ and are obtained using the Vegard's law \cite{vegard1921constitution, gonzalez2007optprop}. In Fig. \ref{fig2}, imaginary part of permittivity are calculated for InAs, GaSb and Ga$_{0.857}$In$_{0.143}$As$_{0.05}$Sb$_{0.95}$ (top cell material of the optimal TPV system) from its dielectric functions. It is clear that the bandgap frequency, sub- and above-bandgap light absorption of the cell material are significantly shifted by variation of $x$ and $y$. Diffusion coefficient ($D_{e,h}$) and the carrier lifetime ($\tau_{e,h}$) of each subcell are calculated from the Caughey-Thomas model \cite{sotoodeh2000JAP} and the Matthiessen's rule $1/\tau=1/\tau_\text{SRH}+1/\tau_\text{Auger}+1/\tau_\text{Radiative}$, respectively. Vegard's law \cite{gonzalez2007optprop} is used again to determine the model parameters and basic properties, such as effective electron/hole mass. Details for obtaining optical and electrical properties of quaternary semiconductor are provided in Section 2 of the \textbf{Supplementary Material}.

The dielectric function of tungsten is obtained from the tabular data in the wavelength range of $\lambda < 10$ $\mu$m \cite{palik1998handbook} and from the LD oscillator model in the range of $\lambda > 10$ $\mu$m \cite{rakic1998optical}. The Drude model is used to determine the dielectric function of Au \cite{ordal1985optical} and ITO \cite{st2017hot, zhao2017high, papadakis2020broadening}. The plasma frequency of ITO $(\omega_p)$ is varied in the range of $0.4-0.9$ eV.

\subsection{Near-field thermal radiation}
When the photon whose energy is greater than the bandgap of the tandem cell is absorbed, an electron-hole pair is generated. If we find the spatial distribution of the steady-state minority carrier concentration in the direction of perpendicular to the TPV cell surface under the illumination condition (i.e., $z$-direction), the photocurrents generated in each subcell can be calculated. In this study, the one-dimensional steady-state continuity equation considering the diffusion and recombination of the minority carrier is considered \cite{vaillon2006modeling, park2008performance, francoeur2011thermal}:
\begin{equation}\label{eq1}
D_{e,h} \frac{d^2 \lbrace n_{e,h}(z, \omega) -n_{e,h}^{0} \rbrace}{dz^2} -  \frac{n_{e,h} (z, \omega)-n_{e,h}^{0}}{\tau_{e,h}} + \dot{g}(z,\omega) = 0
\end{equation}
where $n(z,\omega)$ is the frequency-dependent local carrier concentration, $n^0$ is the equilibrium carrier concentration, and $\dot{g}(z,\omega)$ is the photogeneration rate. The subscript of $e$ and $h$ represents the electron and hole, respectively. In order to calculate the photogeneration rate, one needs to know the spectral near-field radiation absorbed in each subcell at location $z$. When calculating the near-field radiation, each subcell is treated as a homogeneous isotropic medium regardless of the doping type. The net radiative heat flux from the emitter to each tandem TPV cell layer can thus be expressed by
\begin{equation}\label{eq2}
q''_{m}=\int_0^{\infty} d\omega\ q''_{m, \omega} = \int_0^{\infty} d\omega\ \int_0^{\infty} S_{m, \beta, \omega}(\beta, \omega) d\beta
\end{equation}
where $\beta$ is the parallel component of wavevector, and the subscript of $m$ indicates the tandem TPV cell layers from the bottom to the top (e.g., $m=1$ is the Au reflector, $m=2$ is the InAs bottom cell, and $m=5$ is the GaInAsSb top cell). To calculate $S_{m,\beta,\omega}(\beta,\omega)$, we use the standard formalism of fluctuational electrodynamics \cite{polder1971theory, zhang2007nano} as
\begin{equation}\label{eq3}
S_{m, \beta, \omega}(\beta, \omega) =\left \lbrace\Theta(\omega,T_H)-\Theta(\omega,T_L) \right \rbrace \frac{1}{4\pi^{2}} \xi(\omega,\beta) \beta
\end{equation}
where $\Theta(\omega,T)=\frac{\hbar\omega}{\text{exp}\{\hbar\omega/k_BT\}-1}$ is the mean energy of the Planck oscillator with $\hbar$ beting the Planck constant divided by $2\pi$, and $k_B$ is the Boltzmann constant. In Eq. \eqref{eq3}, $\xi(\omega,\beta)$ represents the energy transmission coefficient obtained by employing the formalism considering both forwardly and backwardly propagating (or decaying) waves in each layer \cite{park2008performance, francoeur2009solution}. After we calculate the spectral near-field radiation absorbed in the tandem cell layer by layer, the spatial distribution of the absorbed near-field radiation can be expressed semi-analytically by assuming that each subcell can be treated as a slightly absorbing medium (i.e., $\kappa\ll n$). Notice that in a slightly-absorbing medium, the radiative heat flux can be divided into forward and backward directions with reasonably small error \cite{zhang1997reexamination, lim2019performance}. Therefore,  $S_{m,\beta,\omega}(\beta,\omega)$ can be expressed as 
\begin{equation}\label{eq4}
S_{m, \beta, \omega}(\beta, \omega) = \left \lbrace U_{m, \beta, \omega}(\beta, \omega)+B_{m, \beta, \omega}(\beta, \omega) \right \rbrace \left \lbrace 1-e^{-2\text{Im}(k_{mz})c_{m}} \right \rbrace
\end{equation}
where $U_{m,\beta,\omega}(\beta,\omega)$ and $B_{m,\beta,\omega}(\beta,\omega)$ are the contribution of radiative heat flux by waves in $z$-direction and $-z$-direction, respectively (see Fig.\ \ref{fig1}). In the above equation, $k_{mz}$ presents the normal wavevector component in the $m^\text{th}$ medium, $r_{m2}$ is the Fresnel reflection coefficient at the bottom interface in the $m^\text{th}$ medium, and $c_m$ refers to the total thickness of the $m^\text{th}$ medium. Considering that $B_{m,\beta,\omega}(\beta,\omega) =U_{m,\beta,\omega}(\beta,\omega)e^{-2\text{Im}(k_{mz})c_{m}}\lvert r_{m2} \rvert^{2}$, both $U_{m,\beta,\omega}(\beta,\omega)$ and $B_{m,\beta,\omega}(\beta,\omega)$ can be calculated using Eqs.\ (\ref{eq3}) and (\ref{eq4}). The spectral radiative heat flux estimated with the assumption of slightly absorbing medium is validated in Section 1 of the \textbf{Supplementary Material}.

\subsection{TPV performance}
In this section, we will concentrate on how to obtain the photocurrent generated in the InAs bottom cell and thus omit the subscript of $m$ for simplicity. We define $a=t_{p,\text{bottom}}$, $b=t_{p,\text{bottom}}+ t_{\text{dp,bottom}}$, and $c= t_{p,\text{bottom}}+ t_{\text{dp,bottom}}+ t_{n,\text{bottom}}$, where $t_{p,\text{bottom}}$, $t_{\text{dp,bottom}}$, and $t_{n,\text{bottom}}$ are the width of $p$-region, depletion region, and $n$-region. The width of the depletion region can be calculated by $t_{\text{dp}}=\sqrt{(2\varepsilon_s/e)(k_BT/e)\text{ln}(N_aN_d/n_i^2)(1/N_a+1/N_d)}$, where $\varepsilon_s$ is the static relative permittivity, $e$ is the electron charge, $N_a$ is the $p$-region doping concentration, $N_d$ is the $n$-region doping concentration, and $n_i$ is the intrinsic concentration. The spectral radiation absorbed at location $z$ of the InAs bottom cell can be expressed by
\begin{equation}\label{eq5}
\begin{aligned}
 Q(z,\omega)&
 =\int_0^{\infty} \left [ -\frac{d\{U_{\beta,\omega}(\beta,\omega)e^{-2\text{Im}(k_{2z})z}\}}{dz}+\frac{d\{B_{\beta,\omega}(\beta,\omega)e^{-2\text{Im}(k_{2z})(c-z)}\}}{dz} \right ] d\beta\\ &
 = \int_0^{\infty} \left [ 2\text{Im}(k_{2z})\ U_{\beta,\omega}(\beta, \omega)e^{-2\text{Im}(k_{2z})z} +2\text{Im}(k_{2z})\ B_{\beta,\omega}(\beta, \omega)e^{-2\text{Im}(k_{2z})(c-z)} \right ] d\beta
\end{aligned}
\end{equation}
Then, the photogeneration rate can be calculated by
\begin{equation}\label{eq6}
\dot{g}(z,\omega)=\frac{1}{\hbar\omega}Q(z,\omega)\ \text{ for }\ \omega > \omega_{g}
\end{equation}
where $\omega_g$ indicates the angular frequency corresponding to the bandgap energy of the InAs bottom cell. Using Eq. \eqref{eq6}, the general solution of Eq.\ \eqref{eq1} can be semi-analytically expressed as \cite{lim2015graphene, lim2019performance}:
\begin{equation}\label{eq7}
\begin{aligned}
n_{e,h}(z, \omega) - n_{e,h}^{0}& 
= H_{e,h} \exp \Big ( {\frac{z}{\sqrt{D_{e,h} \tau_{e,h}}}} \Big)
+K_{e,h} \exp \Big( {\frac{-z}{\sqrt{D_{e,h} \tau_{e,h}}}} \Big)\\ &
+ \int_0^{\infty}  \frac{\tau_{e,h} 2\text{Im}(k_{2z})\lbrace U_{\beta, \omega}(\beta,\omega)e^{-2\text{Im}(k_{2z})z} + B_{\beta, \omega}(\beta, \omega)e^{-2\text{Im}(k_{2z})(c-z)}\rbrace}{\hbar\omega\lbrace1- 4D_{e,h} \tau_{e,h} \text{Im}(k_{2z})^2\rbrace} \ d\beta
\end{aligned}
\end{equation}
In the above equation, $H_{e,h}$ and $K_{e,h}$ can be obtained from the following boundary conditions: (i) at the edge of the $p$-doped region, the electron is recombined with the surface recombination velocity $S_e$ $\Big(\text{i.e., }D_{e} \frac{d\{n_{e}(z,\omega)-n_{e}^{0}\}}{dz}\Big \vert_{z=0}\ = S_e\{n_{e}(0,\omega)-n_{e}^{0}\}\Big)$; (ii) the electron and hole concentrations are in equilibrium at the edge of the depletion region (i.e., $n_e=n_e^0$ at $z=a$ and $n_h=n_h^0$ at $z=b$); and (iii) at the edge of the $n$-doped region, the hole is recombined with the surface recombination velocity $S_h$ $\Big(\text{i.e., }D_{h} \frac{d\{n_{h}(z,\omega)-n_{h}^{0}\}}{dz}\Big \vert_{z=c}\ = -S_h\{n_{h}(c,\omega)-n_{h}^{0}\}\Big)$. Referring to recently reported value \cite{kret2021investigation}, the surface recombination velocity at subcell/Au and vacuum/subcell interfaces is set to be 100 m/s and at subcell/tunnel junction interfaces is set to be 10 m/s. The photocurrent density generated by diffusion of the minority carrier concentration in the $p$-doped and the $n$-doped region can be calculated respectively \cite{vaillon2006modeling, park2008performance, francoeur2011thermal}
\begin{equation}\label{eq8}
J_e(\omega) = -eD_{e} \frac{dn_{e}(z,\omega)}{dz}\Big \vert_{z=a}\ \text{ and }\
J_h(\omega) = eD_{h} \frac{dn_{h}(z,\omega)}{dz}\Big \vert_{z=b}
\end{equation}
The drift photocurrent density originating from the swept minority carriers by the built-in electric field is generated in the depletion region. We assume that the drift photocurrent is produced without recombination loss. Thus, it can be written by
\cite{vaillon2006modeling, park2008performance, francoeur2011thermal}
\begin{equation}\label{eq9}
J_{\text{dp}}(\omega) = \frac{e}{\hbar\omega}\int_0^{\infty} \left [ U_{\beta, \omega} ( \beta, \omega) \left \lbrace e^{-2\text{Im}(k_{2z})a}-e^{-2\text{Im}(k_{2z})b} \right \rbrace 
+ B_{\beta, \omega} ( \beta, \omega) \left \lbrace e^{-2\text{Im}(k_{2z})(c-b)}-e^{-2\text{Im}(k_{2z})(c-a)} \right \rbrace \right ] d\beta
\end{equation}
Finally, the total spectral photocurrent density generation is obtained by summing the diffusion and drift photocurrents as:
\begin{equation}\label{eq10}
\begin{aligned}
J_{\text{ph}}(\omega) = | J_e(\omega) | + | J_h(\omega) | + |J_{\text{dp}}(\omega)|
\end{aligned}
\end{equation}
The calculation process described in Eqs.\ \eqref{eq5}--\eqref{eq10} can also be applied to the GaInAsSb top cell (i.e., $m=5$) with proper electrical properties. The spectral photocurrent density calculated by the semi-analytic method is validated in Section 1 of the \textbf{Supplementary Material}. It is worthwhile to mention that it takes only about 1 minute to calculate the spectral photocurrent with the semi-analytic method, while the finite difference method requires around 5 hours using the same computational setup. 

\begin{table}[!b]
\caption{\label{tab1}Bounds for the design variables and the optimal configuration at the 100-nm vacuum gap.}
\footnotesize
\centering
\begin{tabular}{c | c c c c} \hline
 Variable & Lower bound & Upper bound & Interval & Optimal configuration \\
\hline
$t_{p,\text{bottom}}$ (nm) & 100 & 2000 & 10 & 370\\
$t_{n,\text{bottom}}$ (nm) & 100 & 2000 & 10 & 130\\
$t_{p,\text{top}}$ (nm) & 100 & 2000 & 10 & 110\\
$t_{n,\text{top}}$ (nm) & 100 & 2000 & 10 & 480\\
$\omega_p$ (eV) & 0.4 & 0.9 & 0.05 & 0.75\\
$t_\text{ITO}$ (nm) & 0 & 200 & 5 & 50\\
$[x,y]$ & - & - & - & [0.857,0.05]\\
\hline
\end{tabular}
\end{table}

When subcells are monolithic series interconnected, the total voltage can be determined by the summation of voltage drops in each subcell. Accordingly, the total current is limited to the smallest current flowing through each subcell \cite{philipps2018high}. The current-voltage $(J$-$V)$ characteristics of the tandem cell under the illumination condition can be expressed by
\begin{equation}\label{eq11}
V=\frac{k_BT}{e}\text{ln}\Big\{\Big (\frac{J_{\text{ph},1}-J}{J_{0,1}}+1\Big )\Big (\frac{J_{\text{ph},2}-J}{J_{0,2}}+1\Big )\Big\}
\end{equation}
where $J_{\text{ph},k} = \int_{\omega_{g,k}}^{\infty}J_{\text{ph},k}(\omega)d\omega$ is a total photocurrent density with $k=1$ (or 2) indicating the bottom cell and top cell, respectively. The dark saturation current density ($J_{0,k}$) represents the current flowing by diffusion of the carriers at the dark condition and can be obtained by Eq.\ \eqref{eq1} when $\dot{g}(z,\omega)=0$. The boundary conditions are the same with the illumination condition except the electron and hole concentrations at the edge of the depletion region; that is, $n_e-n_e^0=n_e^0\{\text{exp}(eV/k_BT)-1\}$ at $z=a$ and $ n_h-n_h^0=n_h^0\{\text{exp}(eV/k_BT)-1\}$ at $z=b$. Unlike the photocurrent, the dark current is frequency independent and can be written by \cite{francoeur2011thermal, tong2015thin, mckelvey2018solid}:
\begin{equation}\label{eq12}
\begin{aligned}
J_{0} = & \frac{eD_en^2_i}{L_eN_a} \times \frac{S_e\text{cosh}(a/L_e)+(D_e/L_e)\text{sinh}(a/L_e)}{(D_e/L_e)\text{cosh}(a/L_e)+S_e\text{sinh}(a/L_e)} \\
+ & \frac{eD_hn^2_i}{L_hN_d} \times \frac{S_h\text{cosh}\lbrace(c-b)/L_h\rbrace+(D_h/L_h)\text{sinh}\lbrace(c-b)/L_h\rbrace}{(D_h/L_h)\text{cosh}\lbrace(c-b)/L_h\rbrace+S_h\text{sinh}\lbrace(c-b)/L_h\rbrace}
\end{aligned}
\end{equation}
where $L_{e,h}=\sqrt{D_{e,h}\tau_{e,h}}$ is the diffusion length of the electron and hole. The maximum electrical power output of the tandem TPV system, $P_{E}$ is defined by the maximum product of current density and voltage, i.e., 
\begin{equation}\label{eq13}
P_{E} = \text{max}\{J\times V(J)\}\ \text{ for }\ 0 < J < J_{sc}
\end{equation}
where $J_{sc} = \text{min}(J_{\text{ph},1},J_{\text{ph},2})$. The conversion efficiency, $\eta$ is calculated by $\eta=P_{E}/q''_{net}$, where $q''_{net}$ is the total radiative heat flux absorbed in the tandem TPV cell (i.e., $q''_{net}=\sum_{m=1}^{5} q''_{m}$).

\subsection{Optimization}
As shown in Fig.\ \ref{fig1}, the proposed near-field tandem TPV system comprises various design variables. Accordingly, it is necessary to optimize the system configuration to obtain the maximum performance. Since the semi-analytic method provides the significantly reduced calculation time compared to the finite-difference method \cite{lim2015graphene, lim2019performance}, we can readily perform the optimization using the conventional tools, such as the genetic algorithm and the particle swarm optimization. Seven design variables, i.e., thickness of the $p$-doped region $(t_{p,\text{bottom}})$ and $n$-doped region $(t_{n,\text{bottom}})$ of the InAs bottom cell, thickness of the $p$-doped region $(t_{p,\text{top}})$ and $n$-doped region $(t_{n,\text{top}})$ of the GaInAsSb top cell, thickness of the ITO film $(t_\text{ITO})$, plasma frequency $(\omega_p)$ of the ITO, and $[x,y]$ combination of the composition ratio of GaInAsSb top cell, are selected for the optimization. The lower and upper boundaries of design variables are listed in Table \ref{tab1}. The doping concentrations of both $p$- and $n$-region for the bottom cell and top cell are set to be $1\times10^{18}$ cm$^{-3}$, and that of the tunnel junction is assumed to be $1\times10^{19}$ cm$^{-3}$. A genetic algorithm is applied for the objective function that is a conversion efficiency $\eta$. To reduce the computational expenses for the optimization process, we intentionally let six design variables (excluding $[x,y]$ combination) vary by the designated interval (see Table \ref{tab1}). For $[x,y]$ combination, first of all, we found 450 combinations in which the lattice constant of GaInAsSb is within 4.5$\%$ deviation from that of InAs. Then, each combination was designated as a variable and applied to optimization.

\section{Results and discussion}
The optimal configuration of the proposed near-field tandem TPV system is listed in Table \ref{tab1}. The bandgap energy of Ga$_{x}$In$_{1-x}$As$_{y}$Sb$_{1-y}$ quaternary semiconductor is 0.625 eV for the optimal composition ratio of $[x,y]=[0.857,0.05]$. At the vacuum gap of 100 nm, the maximum electrical power output is found to be 8.41 W/cm$^2$ with the conversion efficiency of 35.6\%.

\begin{figure}[!b]
\centering
\includegraphics[width=0.95\textwidth]{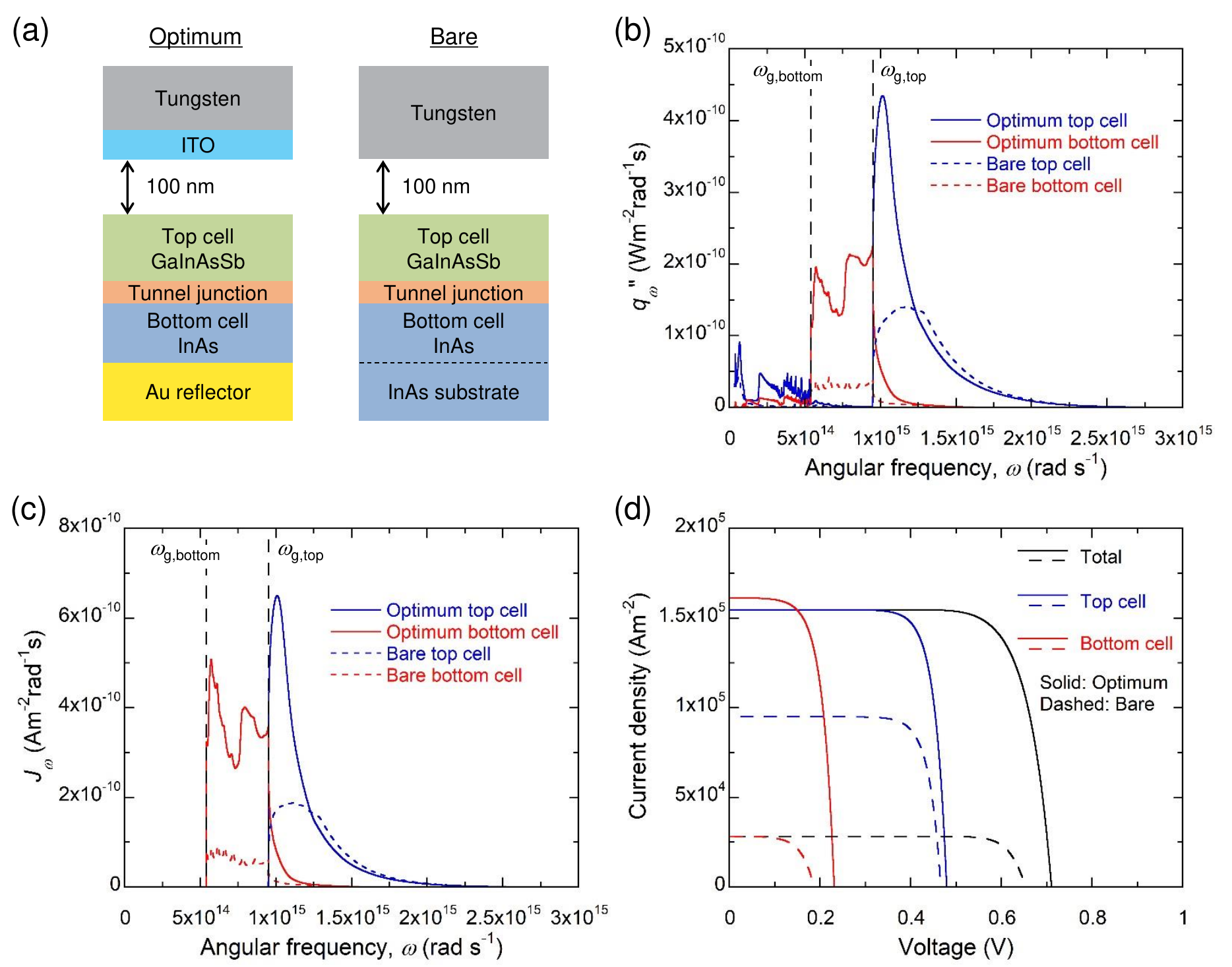}
\caption{\label{fig3}
Optimization results of the near-field tandem thermophotovoltaic system: (a) schematic of the optimized and bare configurations: (b) absorbed spectral radiative heat flux absorbed in the top and the bottom cells; (c) spectral photocurrent density generated in the top and the bottom cells; and (d) current-voltage characteristics under illumination.
}
\end{figure}

To elucidate how the optimal configuration achieves the maximum performance, the absorbed spectral heat flux, the spectral photocurrent, and $J$-$V$ characteristics are analyzed and compared with a bare case. As shown in Fig.\ \ref{fig3}(a), the bare case refers to the configuration without the ITO film and the Au reflector from the optimum case, and all other parameters remain the same. Figure \ref{fig3}(b) compares the spectral radiative heat flux absorbed in the top and the bottom subcells for the optimum and bare cases. It is clear that the spectral radiative heat flux above the bandgap of the top cell is mostly absorbed by the top cell and the sub-bandgap heat flux passing through the top cell is absorbed by the bottom cell. Because the thermalization loss increases as the frequency of absorbed photon shifts far from the bandgap frequency of the TPV cell, the conversion efficiency would decrease if the top cell is absent. By adding a wider-bandgap top cell on the InAs bottom cell, the thermalization loss can be reduced, and the corresponding open-circuit voltage can be increased leading to the performance enhancement. It is also observed from Fig.\ \ref{fig3}(b) that there exists sub-bandgap absorption below $\omega_{g,\text{bottom}}$ for the optimum case while it is negligible for the bare case. In the optimum case, multiple reflections occur due to the Au reflector within the tandem cell, leading to more chances for the sub-bandgap absorption. In particular, the top cell adsorbs more sub-bandgap radiation than the bottom cell given that Ga$_{x}$In$_{1-x}$As$_{y}$Sb$_{1-y}$ has a larger sub-bandgap absorption coefficient than InAs as well as the optimum thickness of the top cell is thicker than the bottom cell. In the bare case, the transmitted radiation through the tandem cell is eventually absorbed by the InAs substrate and is thermally lost. Therefore, the conversion efficiency of the bare case (i.e., 6.8\%) will be much smaller than the optimal case (i.e., 35.6\%.). For quantitative information, the layer-by-layer spectral absorption is provided in Section 3 of the \textbf{Supplementary Material}.

Figure \ref{fig3}(c) describes the spectral photocurrents generated in the top and bottom cells. It can be seen that the absorbed spectral heat flux whose frequency is greater than the bandgap frequency is converted to the spectral photocurrent. Because the sub-bandgap absorption is not substantial for both the top and bottom cells, the spectral photocurrent density looks similar to the absorbed spectral radiative heat flux. On the other hand, in the InAs bottom cell, the spectral photocurrent near $\omega_{g,\text{bottom}}$ becomes noticeable compared to that near $\omega_{g,\text{top}}$ for both the optimum and bare cases. This is because the absorbed spectral heat flux, whose frequency is closer to the bandgap, exhibits less thermalization loss. 

The integrated spectral photocurrent appears as a short-circuit current in the $J$-$V$ characteristics of the TPV cell. In Fig.\ \ref{fig3}(d), the $J$-$V$ characteristics of the near-field tandem TPV cell are presented. Total $J$-$V$ curve shows the total current with respect to the voltage for the series-interconnected tandem cell obtained by Eq.\ \eqref{eq11}. Here, the $J$-$V$ curves of the top and bottom cells indicate characteristics of individual subcell given by $J(V)=J_{\text{ph}}-J_0\{\text{exp}(eV/k_BT)-1\}$. For the bare case, the photocurrent generated in the top cell is much greater than that in the bottom cell. Because the total current of the tandem cell is simply determined by the minimum photocurrent generated in the top and bottom cells, the over-generated photocurrent from the top cell should be thermally lost. This leads to the decrease of the conversion efficiency. Concerning the optimum case, the magnitude of photocurrent produced in each subcell is almost equal. In other words, the current matching condition is achieved for the optimum case. It is interesting to note that the optimization process generates the configuration automatically satisfying the current matching condition just like the high-performance multi-junction cells that are known to satisfy the current matching condition  \cite{hossain2016novel, philipps2018high}.

\begin{figure}[!t]
\centering
\includegraphics[width=0.9\textwidth]{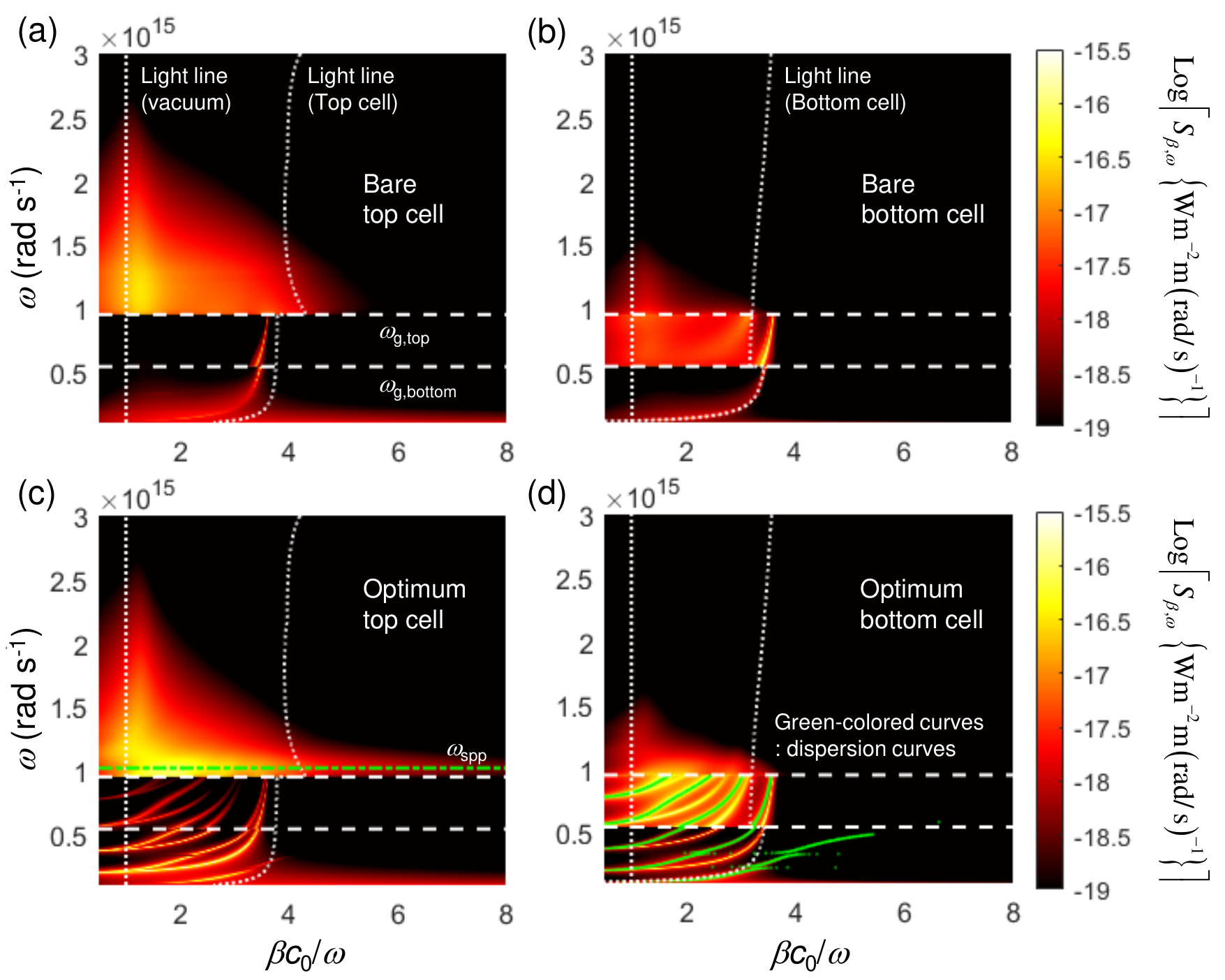}
\caption{\label{fig4}
Contour plots of $S_{\beta,\omega}(\beta,\omega)$ absorbed in (a) the top cell and (b) the bottom cell for the bare case, and (c) the top cell and (d) the bottom cell for the optimum case.
}
\end{figure}

Detailed physical mechanisms for the performance enhancement of the proposed near-field tandem TPV system can be explored by the contour plot of $S_{\beta,\omega}$, as shown in Fig.\ \ref{fig4}. The angular frequency corresponding to the bandgap and the light line of medium $\{\text{i.e., } \beta=n(\omega/c_0)$, where $n$ is the refraction index and $c_0$ is the speed of light in vacuum$\}$ are also displayed. It can be seen from Fig.\ \ref{fig4}(a) that the broad absorption near the thermal characteristic frequency determined by Wien's displacement law for the 1500-K emitter (i.e., $9.75\times 10^{14}$ rad/s).

A large amount of radiative heat is transferred between the vacuum light line and the light line of the top cell, suggesting that near-field radiation through the frustrated mode largely contributes to the absorption in the top cell. As represented in Figs.\ \ref{fig4}(a) and (b), the bare case does not exhibit any resonance behavior. When the tungsten emitter is covered by the thin-ITO film, the SPPs can be supported at the ITO-vacuum interface, and the corresponding resonance frequency ($\omega_{\text{spp}}$) is given by \cite{joulain2005surface}:
\begin{equation}\label{eq15}
\omega_{\text{spp}}=\sqrt{\frac{\varepsilon_\infty}{\varepsilon_\infty+1}\omega_p^2+\Gamma^2}
\end{equation}
where $\varepsilon_\infty=4$ and the damping frequency of ITO, $\Gamma=0.1$ eV \cite{st2017hot, zhao2017high, papadakis2020broadening}. It can be seen from Fig.\ \ref{fig4}(c), the large absorption enhancement occurs near the spectral region corresponding to the SPP resonance frequency indicated as a green dashed line. In fact, the normalized parallel wavevector of $S_{\beta,\omega}$ is extended to a wider range beyond the light line of the top cell, indicating that the surface mode contributes to the absorption in the top cell. Since the thickness of the ITO film is thin compared to the vacuum gap distance, the emission of tungsten still contributes to radiative heat transfer.

Figure \ref{fig4}(d) reveals that there exists additional resonance mode different from the SPP in the bottom cell. It turns out that a trapped waveguide mode exists in a thin TPV cell such that photons are confined inside the TPV cell by decaying fields in tungsten emitter and Au reflector \cite{zhang2007nano, tong2015thin}. The green-colored dispersion curves for the whole NFTPV system are obtained by solving Maxwell’s equation with the proper boundary conditions \cite{yeh2008essence}. We can see that the dispersion curves are overlapped with where the large absorption occurs. As shown in Fig.\ \ref{fig3}(b), the amount of radiation absorbed by the bottom cell is greatly improved by the trapped waveguide mode. As the absorption of the bottom cell is increased, the sub-bandgap absorption in the top cell is also increased by the waveguide mode $\{$see Fig.\ \ref{fig4}(c)$\}$. However, in the frequency region of $\omega$$>$$\omega_{\text{g,top}}$ where the absorption coefficient is large, the waveguide resonance hardly causes additional absorption in the top cell. If the InAs bottom cell becomes thicker, the resulting photocurrent by the bottom cell could be larger with the aid of the increased number of resonance branches due to the trapped waveguide mode. This enhancement would be crucial as the vacuum gap distance becomes smaller because the absorption in the bottom cell should also be increased along with the strong absorption enhancement in the top cell due to the largely improved evanescent mode. 

To improve the conversion efficiency of the near-field tandem TPV system, the absorption of two subcells must be increased together to satisfy the current matching condition. In the proposed system, we exploit two different resonance modes. In the top cell, the surface mode supported by the SPP resonance in the ITO-vacuum single interface is the main physical mechanism of the absorption enhancement. Owing to the trapped waveguide mode formed in the thin TPV cell, the large photocurrent can be generated in the bottom cell. Individual effects of the thin-ITO film and the Au reflector on the radiative heat flux are provided in Section 4 of the \textbf{Supplementary Material}.


\begin{figure}[!t]
\centering
\includegraphics[width=0.9\textwidth]{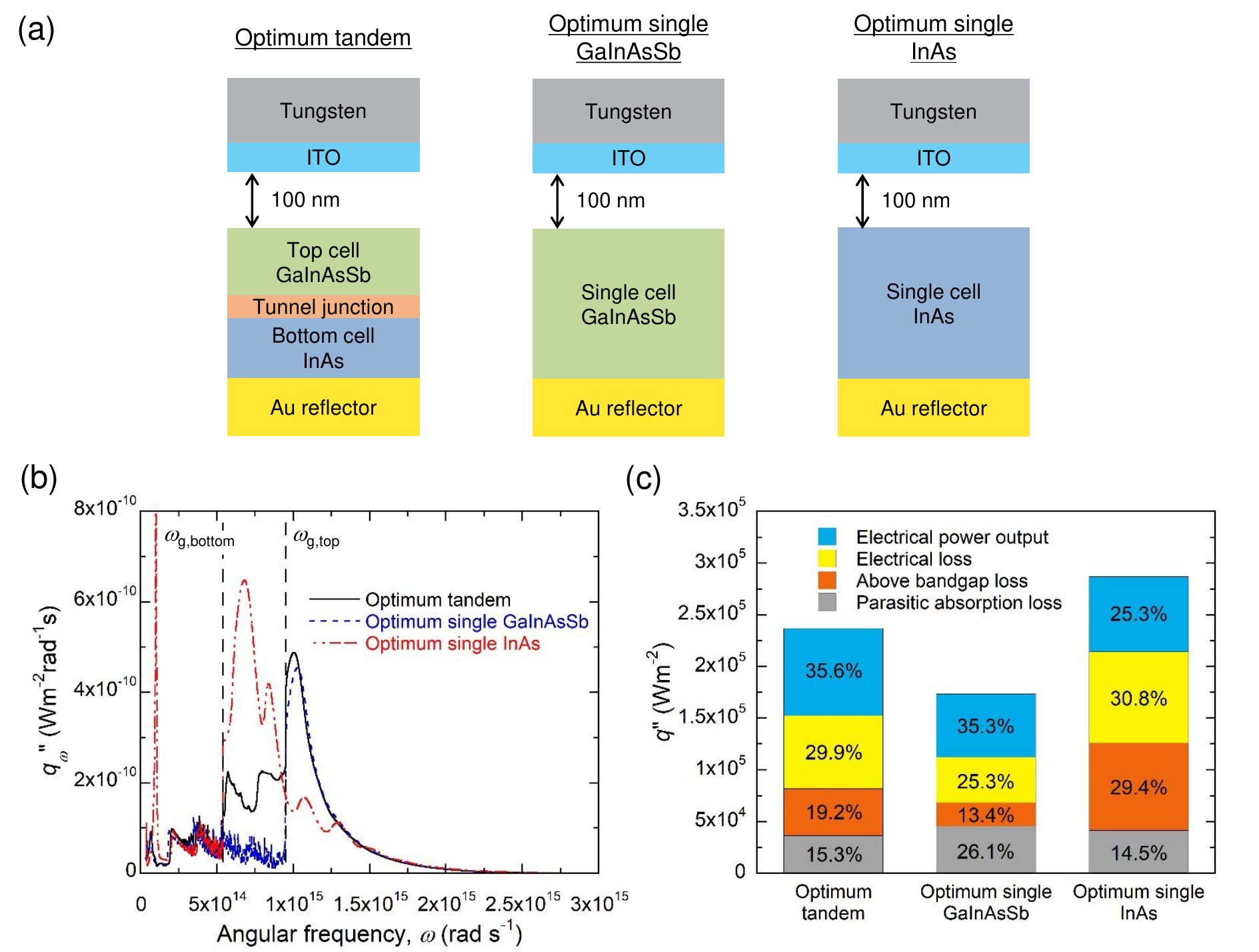}
\caption{\label{fig5}
Comparison of tandem-cell-based and single-cell-based near-field thermophotovoltaic systems: (a) schematic of optimized tandem, single GaInAsSb, and single InAs cases; (b) total spectral radiative heat flux absorbed in the TPV cell; and (c) loss analysis for three cases described in (a).
}
\end{figure}

To clearly demonstrate advantages of the tandem-cell-based NFTPV system compared to conventional single-cell system, two single-cells systems made of GaInAsSb or InAs cell $\{$refer to Fig.\ \ref{fig5}(a)$\}$ are optimized separately. Here, the design variables of the single-cell-based NFTPV system are the thickness of $p$- and $n$-doped region of TPV cell, $[x,y]$ combination of the composition ratio (only for GaInAsSb cell), the thickness of ITO, and the plasma frequency of ITO. The upper and lower boundaries of design variables follow those listed in Table \ref{tab1}. For fair comparison, we impose a constraint such that the total thickness of a single cell is the same as that of the optimum tandem cell $\{$i.e., 1191 nm including the tunnel junction (20 nm) and depletion regions of GaInAsSb (46 nm) and InAs (35 nm) subcells$\}$. The optimal configurations of two single-cell-based NFTPV systems are listed in Table \ref{tab2}.

Figure \ref{fig5}(b) shows the spectral heat flux absorbed in the entire TPV cell, including the tunnel junction and the backside Au reflector. Comparing the optimum tandem case and the optimum single GaInAsSb case, the absorbed radiative heat flux looks identical in the frequency range above $\omega_{g,\text{top}}$. This is due to the similar contribution of the surface mode at the ITO film. On the other hand, the net above-bandgap heat flux absorbed in the single InAs TPV cell is totally different from the absorption in the bottom cell of the optimum tandem case. This spectral difference can be explained by comparing the optimal ITO conditions. Because the conversion efficiency increases as the radiation is absorbed closer to the bandgap frequency, the SPP resonance frequency supported by the ITO-vacuum interface is tuned to be located closer to $\omega_{g,\text{bottom}}$ (i.e., bandgap frequency of InAs) for the single InAs case. In addition, the ITO thickness of the single InAs case is much thicker than that of the optimum tandem case (i.e., 120 nm $>$ 50 nm). This is because when the thermal characteristic frequency determined by Wien's displacement law and the bandgap frequency of the TPV cell are completely mismatched, the broad emission effect arouse by the tungsten should be suppressed and the selective emission by ITO has to be emphasized.

\begin{table}[!b]
\caption{\label{tab2}Optimal configuration of single-cell-based near-field thermophotovoltaic systems.}
\begin{tabular}{c | c c c c c} \hline
  & $t_{p}$ (nm) & $t_{n}$ (nm) & $\omega_p$ (eV) & $t_\text{ITO}$ (nm) & $[x,y]$ \\
\hline
Single GaInAsSb cell & 380 & 770 & 0.75 & 40 & [0.857,0.05]\\
Single InAs cell & 970 & 190 & 0.55 & 120 & -\\
\hline
\end{tabular}
\end{table}

The performance of the tandem-cell-based and single-cell-based NFTPV systems are comprehensively analyzed by dividing the total absorbed radiative heat flux $\{$i.e., integration of the spectral heat flux shown in Fig.\ \ref{fig5}(b)$\}$ into electrical power output, electrical loss, above-bandgap loss, and parasitic absorption loss. Figure \ref{fig5}(c) compares the occupying percentage of the aforementioned components for three cases. The ratio of the electrical power output to the total absorbed radiative heat flux indicates simply the conversion efficiency $\eta$. The electrical loss of the optimum tandem case can be calculated by
\begin{equation}\label{eq16}
P_\text{loss,electrical} = \sum_{m=2,5}E_{g,m}\int_{\omega_{g,m}}^{\infty}\frac{q''_{m,\omega}}{\hbar\omega}d\omega - P_{E}
\end{equation}
where $E_g$ is the bandgap energy of the semiconductor. The first term of the right-hand side of Eq.\ \eqref{eq16} indicates the power output when all photogenerated electron-hole pairs contribute to the generation of the electrical power with the potential corresponding to the bandgap energy. The presence of the dark current and the recombination loss which makes the quantum efficiency less than unity causes the electrical loss. The above-bandgap loss for the optimum case can be calculated by
\begin{equation}\label{eq17}
P_\text{loss,above bandgap} = \sum_{m=2,5}\int_{\omega_{g,m}}^{\infty}\frac{\omega-\omega_g}{\omega}q''_{m,\omega}d\omega
\end{equation}
implying that the difference between the photon energy and the bandgap energy causes the thermalization loss. The electrical and above-bandgap losses for the optimum single cases also can be estimated from Eqs.\ \eqref{eq16} and \eqref{eq17} by considering only one TPV cell. The parasitic absorption loss includes the absorbed radiation in the tunnel junction and the Au reflector, as well as the sub-bandgap absorption.

\begin{figure}[!b]
\centering
\includegraphics[width=0.5\textwidth]{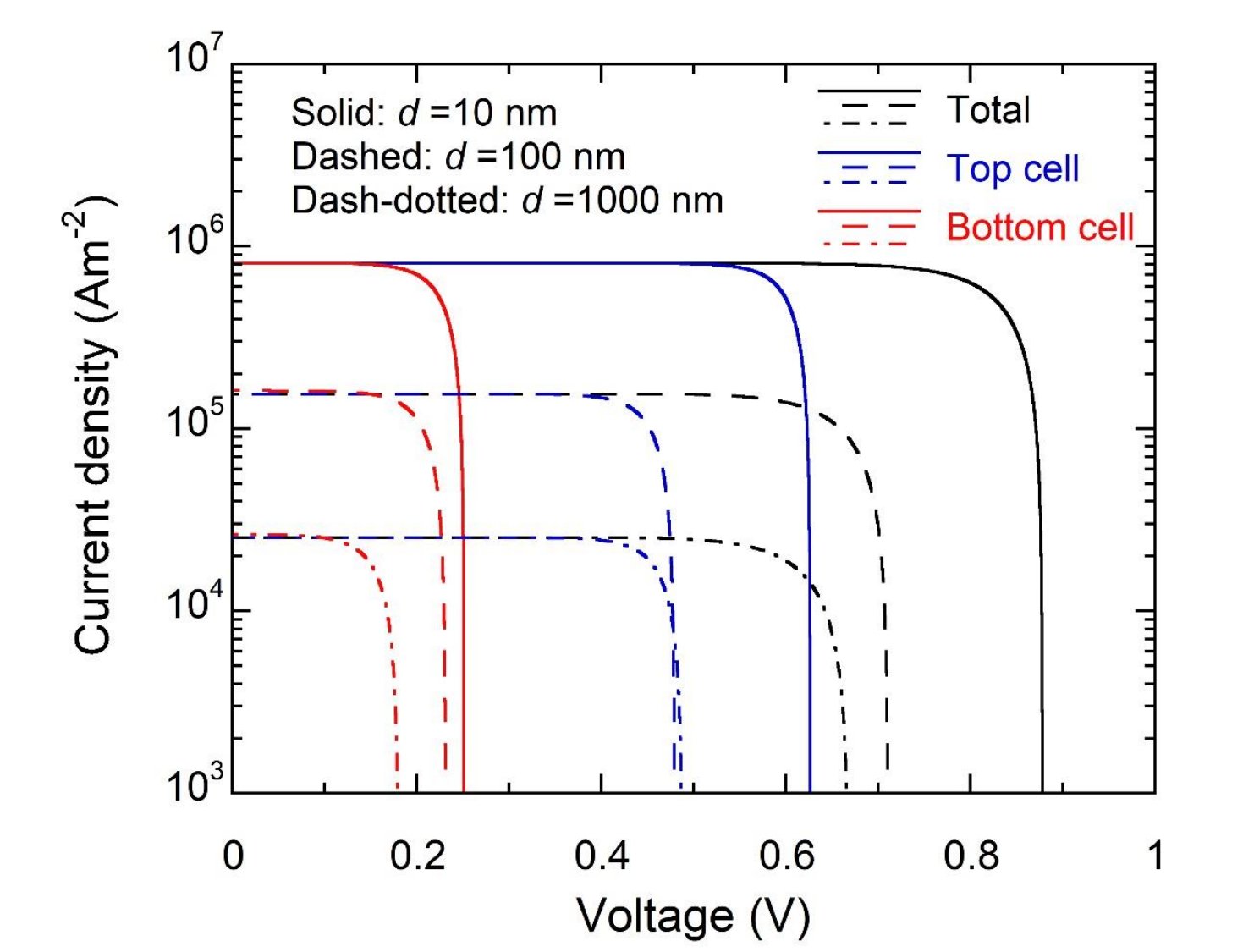}
\caption{\label{fig6}
Current-voltage characteristics under illumination at $d=10$, 100, and 1000 nm.}
\end{figure}

Firstly, the result of the loss analysis of the optimum tandem is compared with that of the optimum single GaInAsSb. Because the dark current of the InAs cell in the tandem TPV system is large, the electrical loss of the optimum tandem case is larger even though the open-circuit voltage of the tandem cell is enhanced satisfying the current matching condition. In addition, while the quasi-monochromatic radiation is absorbed in the top cell of the tandem TPV or single GaInAsSb cell, the bottom cell has multiple spectral peaks due to the trapped waveguide modes, which rather increases the above bandgap loss. However, the slightly higher conversion efficiency is achieved for the optimum case due to the significant advantage in the parasitic absorption loss, particularly in the sub-bandgap absorption. Moreover, the maximum electrical power output generated in the optimum tandem case is 1.4 times larger than the optimum single GaInAsSb case. Therefore, the overall performance of the NFTPV system with the optimum tandem cell is better than that of the optimum single GaInAsSb cell is used. 

The major difference in the loss analysis between the optimum tandem case and the optimum single InAs case is the above-bandgap loss. The optimum single InAs case shows the largest above-bandgap loss even though $\omega_p$ is tuned such that $\omega_{\text{spp}}$ is located near the bandgap frequency of the InAs. Because the total thickness of a single InAs cell is the same as that of a tandem cell, thickness-dependent waveguide modes are excited in a way to improve the above-bandgap absorption of the InAs cell, similarly to the case of tandem cell. In fact, Fig. \ref{fig5}(b) shows the major peak supported by SPPs (near $7.6\times10^{14}$ rad/s) as well as multiple-minor peaks due to waveguide modes for the optimum single InAs case. Because minor peaks deteriorate the quasi-monochromatic behavior, the optimum single InAs case eventually shows a smaller conversion efficiency than the optimum tandem case.


\begin{table}[!t]
\caption{\label{tab3}Optimal configuration of near-field tandem thermophotovoltaic system at 10 and 100 nm vacuum gap distances.}
\begin{tabular}{c| c c c c c c c} \hline
 Vacuum gap & $t_{p,\text{bottom}}$ (nm) & $t_{n,\text{bottom}}$ (nm) & $t_{p,\text{top}}$ (nm) & $t_{n,\text{top}}$ (nm) & $\omega_p$ (eV) & $t_\text{ITO}$ (nm) & $[x,y]$ \\
\hline
$d=10$ nm & 1540 & 100 & 100 & 100 & 0.85 & 10 & [0.943,0.4]\\
$d=1000$ nm & 520 & 150 & 460 & 300 & 0.75 & 50 & [0.921,0.3]\\
\hline
\end{tabular}
\end{table}

Finally, the NFTPV system configuration is again optimized at different vacuum gap distances: 10 nm and 1000 nm. When the vacuum gap ($d$) between the emitter and TPV cell is 10 nm, the radiative heat transfer is significantly enhanced with contribution of the surface mode. At $d=1000$ nm, however, only a marginal amount of the evanescent wave can contribute to the radiative heat transfer. Hence, the propagative mode dominates the radiative heat transfer at $d=1000$ nm. It is interesting to note that although the heat transfer mechanism is changed depending on the vacuum gap distance, current matching conditions can be achieved for three different vacuum gap conditions (see Fig.\ \ref{fig6}). Their optimal configurations are listed in Table \ref{tab3}. 
\begin{figure}[!b]
\centering
\includegraphics[width=0.8\textwidth]{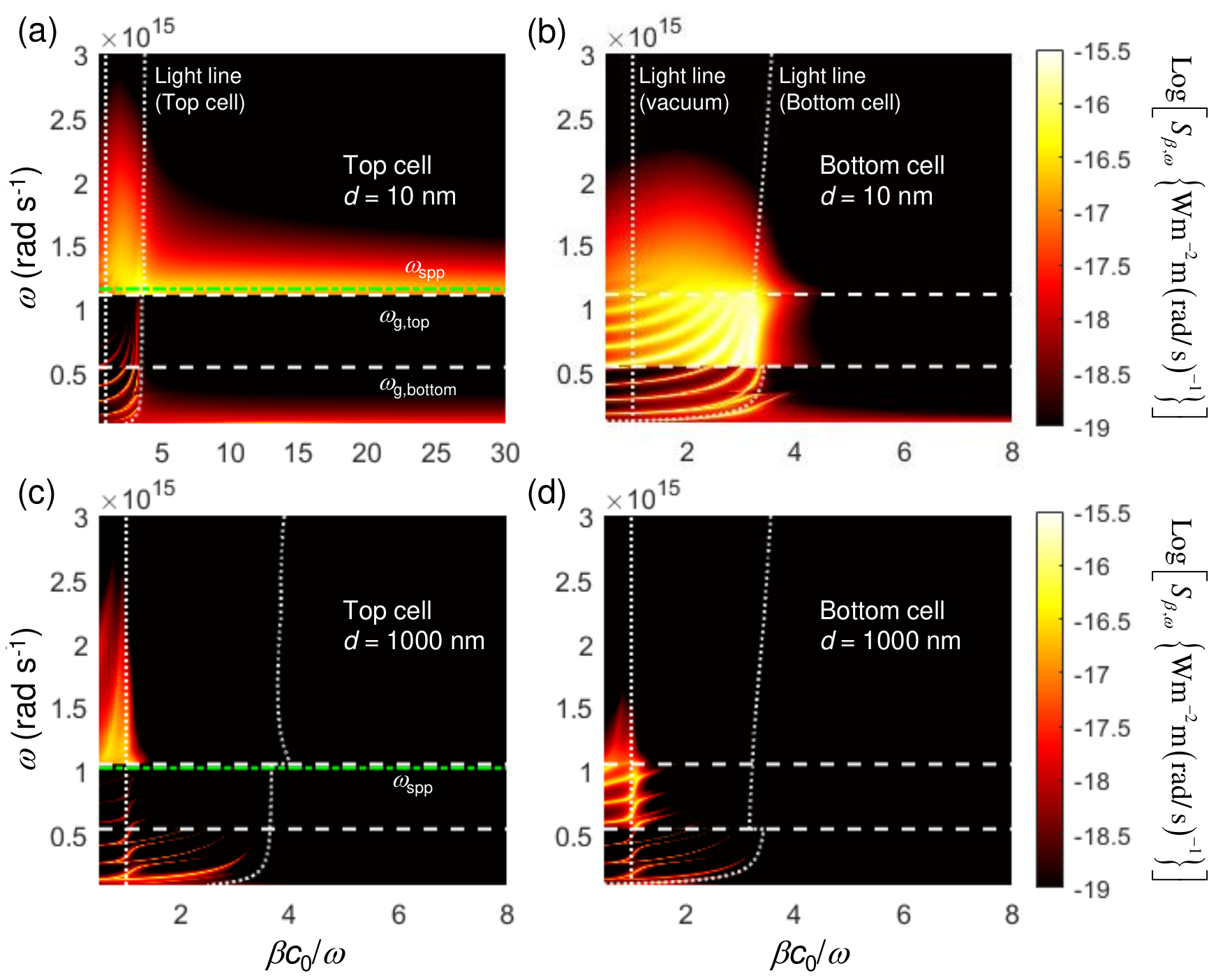}
\caption{\label{fig7}
Contour plots of $S_{\beta,\omega}(\beta,\omega)$ absorbed in (a) top cell at $d=10$ nm, (b) bottom cell at $d=10$ nm, (c) top cell at $d=1000$ nm, and (d) bottom cell at $d=1000$ nm.
}
\end{figure}

Contours of $S_{\beta,\omega}$ absorbed in the top and bottom cells are plotted in Fig.\ \ref{fig7} (for vacuum gaps of 10 nm and 1000 nm) to explain how the current matching condition could be reached at different vacuum gap distances. In Fig.\ \ref{fig7}(a), the normalized parallel wavevector range (i.e., $x$-axis) is extended to 30 to demonstrate the large absorption by the surface mode. Although at $d=10$ nm, the optimum thickness of the top cell is converged to the lower bound (refer to Table \ref{tab1}), the top cell can absorb a large amount of the radiation given that a short penetration depth attributed to a large parallel wavevector \cite{basu2009ultrasmall}. Because of the strong emission through the surface mode, design variables of ITO are optimized in a way to suppress the absorption in the top cell to approach the current matching condition. On the other hand, the total optimum thickness of the tandem TPV cell at $d=10$ nm shows the largest value among three vacuum gap cases. The number of resonance branches formed by the trapped waveguide mode can be increased as the thickness of the TPV cell increases $\{$i.e., comparing Fig.\ \ref{fig7}(b) with Fig.\ \ref{fig4}(d)$\}$. Because those resonance branches boost the absorption, the photocurrent generated in the bottom cell can follow that generated in the top cell (i.e., current matching condition). When at $d=1000$ nm, the absorption in the tandem cell is dominated by the propagative mode where the normalized parallel wavevector is smaller than the vacuum light line $\{$see Figs.\ \ref{fig7}(c) and \ref{fig7}(d)$\}$. Because the radiation propagating in the tandem cell dominates the heat transfer both at $d=100$ nm (i.e., via frustrated mode) and at $d=1000$ nm (i.e., via propagative mode), the optimal conditions of the ITO and the thickness of the tandem cell are similar to each other. As shown in Fig.\ \ref{fig7}(d), even at the large vacuum gap distance of 1000 nm, the trapped waveguide mode is still a main physical mechanism to increase the absorption of the bottom cell.

\section{Conclusions}
We have analyzed a tandem-TPV-cell-based NFTPV system consisting of a thin-ITO-covered tungsten emitter and a GaInAsSb/InAs monolithic interconnected tandem TPV cell. To develop the simulation model of the proposed NFTPV system, firstly, the spatial distribution of absorbed near-field radiative heat flux was attained by separating the forward and backward waves in each subcell. We then calculated the spectral and spatial photocurrent generated in each subcell using a semi-analytically solved minority carrier’s concentration distribution. With the genetic algorithm, the conversion efficiency of 35.6$\%$ and the electrical power output of 8.41 W/cm$^2$ were achieved at $d=100$ nm when the emitter and TPV cell are at 1500 K and 300 K, respectively. The optimal configuration of the NFTPV system satisfied the current matching of subcells, which is an essential design feature of achieving the high-performance multi-junction cells. We elucidated that the physical mechanism of the performance enhancement originates from two resonance modes, i.e., SPP resonance supported by the ITO-vacuum interface and waveguide mode excited in the thin tandem cell. Based on the loss analysis, it was found that the overall performance, including the conversion efficiency and the electrical power output of the tandem-cell-based NFTPV system, is superior compared to the single-cell-based NFTPV system. In addition, at far-to-near field vacuum gap distances, the proper utilization of two resonance modes can yield high-performance near-field tandem TPV systems, achieving current matching conditions. Since the simulation model developed in this study can be applied to emitters using multilayers or HMMs other than thin films, this work will facilitate the development of the further enhanced NFTPV system in the future.

\printcredits

\section*{Declaration of competing interest}
The authors declare that they have no known competing financial interests or personal relationships that could have appeared to influence the work reported in this paper.

\section*{Acknowledgements}
This research is supported by the Basic Science Research Program (NRF-2019R1A2C2003605) through the National Research Foundation of Korea (NRF) funded by Ministry of Science and ICT.

\bibliographystyle{elsarticle-num-names}
\bibliography{Song_Bib.bib}

\end{document}


\setstretch{1.5}
\section*{Supplementary material:\\
Comprehensive analysis of optimized near-field tandem thermophotovoltaic system}

\textbf{Authors}: Jaeman Song$^{1,2}$, Minwoo Choi$^{1,2}$, Mikyung Lim$^{1,2}$, Jungchul Lee$^{1,2}$, Bong Jae Lee$^{1,2*}$\\

\textit{1. Department of Mechanical Engineering, Korea Advanced Institute of Science and Technology, Daejeon 34141, South Korea\\
2. Center for Extreme Thermal Physics and Manufacturing, Korea Advanced Institute of Science and Technology, Daejeon 34141, South Korea\\
3. Nano-Convergence Mechanical Systems Research Division, Korea Institute of Machinery and Materials, Daejeon 34103, South Korea
}\\

*Corresponding authors\\
*e-mail: bongjae.lee@kaist.ac.kr (Bong Jae Lee)\\

\setstretch{1.5}

\newpage
\section*{1. Theoretical model validation}
To validate the spectral radiative heat flux calculated by the energy flow split and the spectral photocurrent density simulated by the semi-analytic method, a near-field TPV system consisting of a GaSb/InAs tandem TPV cell and the tungsten emitter is introduced (see Fig.\ S1). Two subcells are connected by a tunnel junction and the total thickness of the tandem TPV cell is limited by the Au backside reflector. The vacuum gap distance between the emitter and the TPV cell is set to be 100 nm and temperatures of the emitter and the TPV cell are 1500 and 300 K, respectively. In order to validate the simulation model regardless of the TPV cell thickness, we calculate the spectral radiative heat flux absorbed by each subcell and the generated spectral photocurrent density for thick subcells ($t_p=t_n=500$ nm) and thin subcells ($t_p=t_n=100$ nm). In Figs.\ S2(a),(b), the spectral radiative heat flux solved by the original way $\{$Eq.\ (3) of the manuscript$\}$ and by splitting the energy flow $\{$Eq.\ (4) of the manuscript$\}$ are described as lines and symbols, respectively. For thick subcells $\{$Fig.\ S2(a)$\}$ and thin subcells $\{$Fig.\ S2(b)$\}$, we can see that the radiation results are exactly the same. Next, to verify the simulation of the spectral photocurrent density, we compared the results obtained by the finite different method (FDM) and by the semi-analytic method. In Figs.\ S2(c),(d), it can be confirmed that the photocurrent calculated by the semi-analytic method and FDM almost coincides regardless of the thickness of the tandem TPV cell (within 0.4$\%$ error).

\begin{figure}[b!]
\centering\includegraphics[width=10cm]{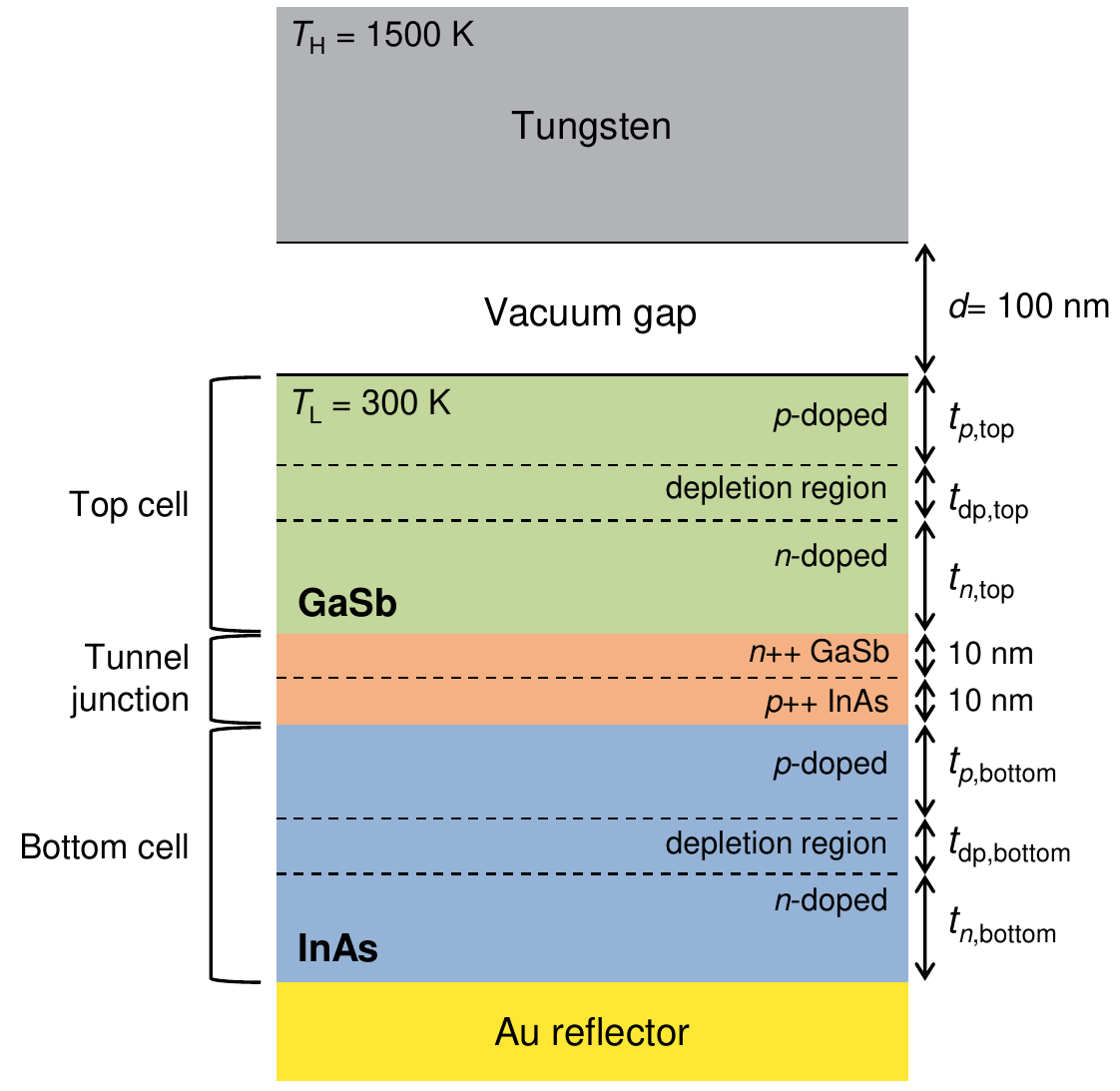}
\caption{\label{Figs1}Fig.\ S1: Schematic illustration of near-field tandem thermophotovoltaic system for the model validation.}
\end{figure}

\newpage
\begin{figure}[h]
\centering\includegraphics[width=17cm]{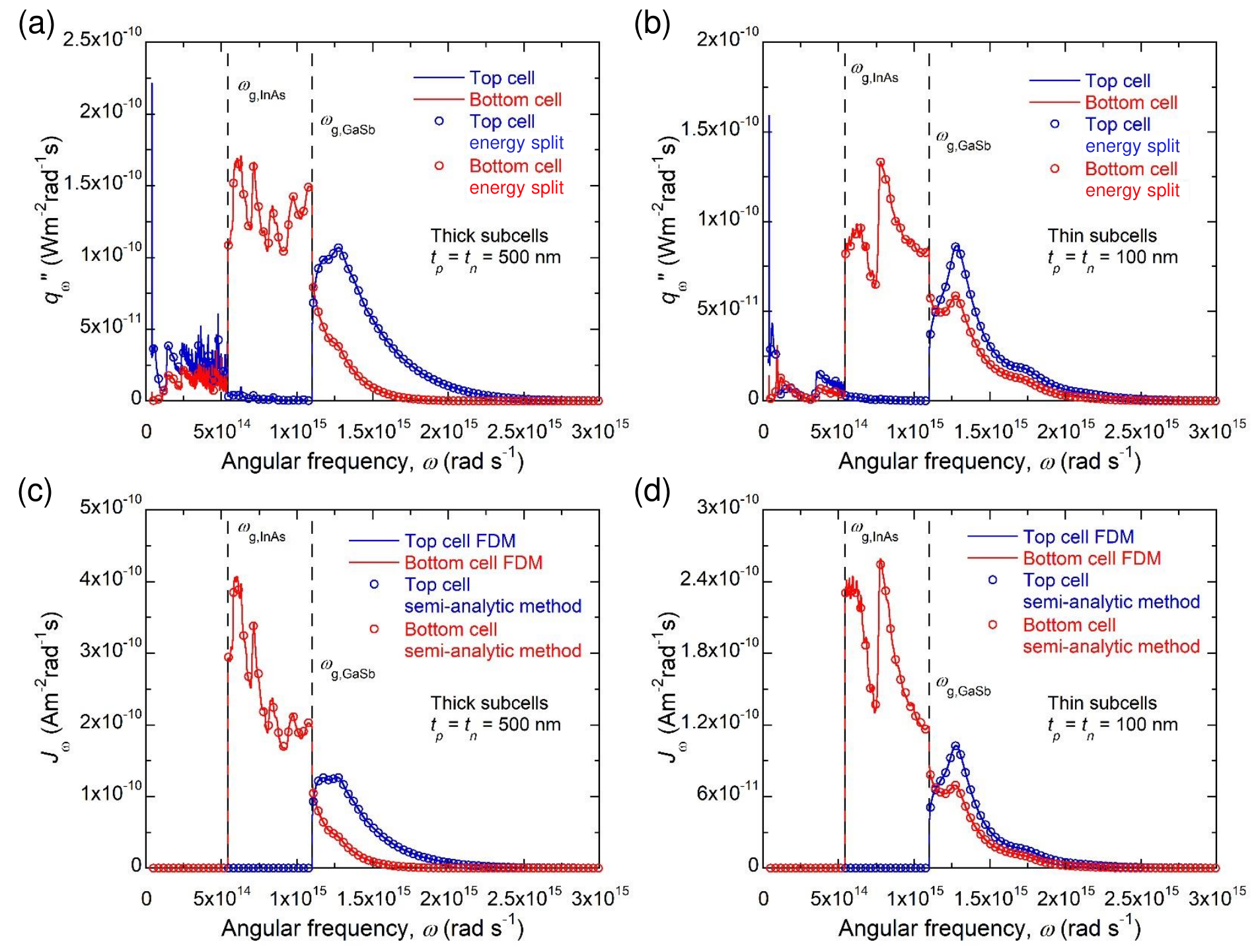}
\caption{\label{Figs2}Fig.\ S2: Theoretical model validation. Spectral radiative heat flux (a) for thick subcells and (b) for thin subcells. Spectral photocurrent density (c) for thick subcells and (d) for thin subcells.}
\end{figure}

\newpage
\section*{2. Optical and electrical properties of semiconductors}
In order to obtain the material property of quaternary solid solution Ga$_{x}$In$_{1-x}$As$_{y}$Sb$_{1-y}$ with arbitrary composition $x, y$ $\in$ [0,1], Vegard's law is extensively utilized. Material property $P$ of a ternary compound A$_{x}$B$_{1-x}$C can be obtained by simple linear combination of two binaries AB and AC accounting for the composition ratio $x$: $P_\text{ABC} = P_\text{AB}+xP_\text{AC}+x(1-x)W_\text{ABC}$. $W_\text{ABC}$ is a correction factor called bowing parameter, determined by measurement data. Furthermore, material property of a quaternary compound A$_{x}$B$_{1-x}$C$_{y}$D$_{1-y}$ can be determined either by Eq.\ \eqref{eq1'} or Eq.\ \eqref{eq2'}, depending on the bowing parameter data available \cite{adachi1987band}.

\begin{equation}\label{eq1'}
P_{\text{ABCD}}=\frac{x(1-x)\{yP_{\text{ABC}}+(1-y)P_{\text{ABD}}\}+y(1-y)\{xP_{\text{ACD}}+(1-x)P_{\text{BCD}}\}}{x(1-x)+y(1-y)}
\end{equation}

\begin{equation}\label{eq2'}
P_{\text{ABCD}}=xyP_{\text{AC}}+x(1-y)P_{\text{AD}}+(1-x)yP_{\text{BC}}+(1-x)(1-y)P_{\text{BD}}\\+x(1-x)W_{\text{AB-}}+y(1-y)W_{\text{-CD}}
\end{equation}

Eq.\ \eqref{eq1'} can be applied only when bowing parameters for all four ternaries W$_\text{ABC}$, W$_\text{ABD}$, W$_\text{ACD}$, W$_\text{BCD}$ are known. If one parameter W$_\text{ABD}$ is unknown, Eq.\ \eqref{eq2'} is used instead approximating W$_\text{ABC}$ as W$_\text{AB-}$ and letting an average value of W$_\text{ACD}$ and W$_\text{BCD}$ as W$_\text{-CD}$. Occasionally, no bowing parameter data may be available. Vegard's law may be directly applied to the material property of interest (P = $\epsilon$, $\mu$, $\tau$) but we apply it to the parameters and coefficients in an appropriate model describing the properties for better physical representation.

In order to model Dielectric function $\epsilon(x,y,\omega)$ of a Ga$_{x}$In$_{1-x}$As$_{y}$Sb$_{1-y}$ semiconductor above its bandgap energy $E_{0}$, Adachi's semi-empirical model \cite{adachi1987band} that accounts for various energy-level transitions described in the material's energy-band structure is used. Direct transition occurring at the center of Brillouin zone ($E_{0}$), three types of off-centered direct transitions ($E_{1}, E_{1}+\Delta_{1}, E_{2}$) and indirect transition ($E_{ID}$) are considered as major contributions. They are separately modeled into corresponding dielectric functions $\epsilon_{<E_{0},E_{1},E_{1}+\Delta_{1},E_{2},E_{ID}>}$ and its weighted summation with strength parameters gives dielectric function as Eq.\ \eqref{eq3'}.
\begin{equation}\label{eq3'}
\epsilon(\omega)=a_{1}\epsilon_{E_{0}}(\omega)+b_{1}\epsilon_{E_{1}}(\omega)+b_{2}\epsilon_{E_{1}+\Delta_{1}}(\omega)+c_{1}\epsilon_{E_{2}}(\omega)+d_{1}\epsilon_{E_{ID}}(\omega)
\end{equation}
First, Temperature-dependent transition energies $E_0, E_1, E_{1}+\Delta_{1}, E_{2}$ are driven for four binaries GaAs, GaSb, InAs, InSb using Varshini's equation $E(T)=E(T=0)-\frac{\delta{T}^2}{T+\beta}$. With Eq.\ \eqref{eq1'}, four transition energies and five strength parameters are merged to find its value for Ga$_{x}$In$_{1-x}$As$_{y}$Sb$_{1-y}$ with composition ratio $[x,y]$ in desire. From now on, subscript $Q$ states for quaternary compound. After $E_{0,\text{Q}}$, $\cdot \cdot \cdot$ , $E_{ID,\text{Q}}$ and $a_{1,\text{Q}}$, $\cdot \cdot \cdot$ , $d_{1,\text{Q}}$ are derived, Eq.\ \eqref{eq3'} results in dielectric function of GaInAsSb above its bandgap energy.  Varshini's equation coefficients $\delta$ and $\beta$, detailed expressions for function $\epsilon_{<E_{0},E_{1},E_{1}+\Delta_{1},E_{2},E_{ID}>}$, Binary compound properties and bowing parameters are given in work of Gonzalez-Cuevas et al. \cite{gonzalez2007optprop}.

We conduct similar procedure to obtain electrical properties $\mu_{e}$, $\mu_{h}$, $\tau_{e}$, and $\tau_{h}$ of Ga$_{x}$In$_{1-x}$As$_{y}$Sb$_{1-y}$. Caughey-Thomas empirical model expressed as Eq.\ \eqref{eq4'} describes electron and hole mobility $\mu_{e}$ and $\mu_{h}$ \cite{caughey1967carrier}.
\begin{equation}\label{eq4'}
\begin{split}
\mu(N_{<A,D>},T)_{<e,h>} = \mu_{\text{min},<e,h>} + (\mu_{\text{max},<e,h>}{(\frac{300}{T})}^{\theta_{1,<e,h>}}-\mu_{\text{min},<e,h>})\\\times\Big({1+{(\frac{N_{A,D}}{N_{\text{ref},<e,h>}{(\frac{T}{300})}^{\theta_{2,<e,h>}}})}^{\Phi_{<e,h>}}}\Big)^{-1}
\end{split}
\end{equation}
In this work, Temperature is $T=300K$, Doping concentrations are $N_{A,D}=10^{18} {\text{ cm}}^{-3}$ for two subcells and $10^{19} {\text{ cm}}^{-3}$ for tunnel junction. Temperature coefficients $\theta_{1,<e,h>}$ and $\theta_{2,<e,h>}$, mobility at large and small doping levels $\mu_{\text{max},<e,h>}$ and $\mu_{\text{min},<e,h>}$, fitting parameter $\Phi_{<e,h>}$, reference doping concentration $N_{\text{ref},<A,D>}$ at which mobility becomes half of $\mu_{\text{max},<e,h>}$, need to be found. With values for binaries GaAs \cite{sotoodeh2000empirical}, InAs \cite{sotoodeh2000empirical}, GaSb \cite{gonzalez2006modeling} and InSb \cite{gonzalez2006modeling}, Vegard's law used again to find values for Ga$_{x}$In$_{1-x}$As$_{y}$Sb$_{1-y}$. Since there are little information about the bowing effect of these parameters, Eq.\ \eqref{eq2'} is used instead of Eq.\ \eqref{eq1'} with zero bowing parameters. Substituting the obtained $\theta_{1,<e,h>,Q}$, $\theta_{2,<e,h>,Q}$, $\mu_{\text{max},<e,h>,Q}$, $\mu_{\text{min},<e,h>,Q}$, $\Phi_{<e,h>,Q}$, and $N_{\text{ref},<A,D>,Q}$ into Eq.\ \eqref{eq4'} gives carrier mobility $\mu_{<e,h>,Q}$.

In order to obtain carrier lifetime $\tau_{<e,h>}$, effect of radiative recombination, Shockley-Read-Hall(SRH) trap-assisted recombination \cite{hall1952electron, shockley1952statistics}, Auger recombination are combined using Matthiessen's rule.
\begin{equation}\label{eq5'}
\begin{split}
{\tau_{<e,h>}}^{-1}={\tau_{\text{rad},<e,h>}}^{-1}+{\tau_{\text{SRH},<e,h>}}^{-1}+{\tau_{\text{Auger},<e,h>}}^{-1}\\={(\frac{1}{B_{\text{rad}}N_{<A,D>}})}^{-1}+{(\frac{1}{{\sigma}{N_t}}\sqrt{\frac{m_{<e,h>}}{3k_BT}})}^{-1}+{(\frac{1}{C_{\text{Auger}}{n_i}^2})}^{-1},\\\text{where } {n_i}^2={N_c}{N_v}{\exp{\frac{-eE_{0}}{k_BT}}} \text{ and } N_{<c,v>}=2{(\frac{2{\pi}m_{e,h}k_BT}{h^2})}^{\frac{3}{2}}
\end{split}
\end{equation}

Intrinsic carrier concentration $n_{i,Q}$ is calculated from third row of Eq.\ \eqref{eq5'} after finding $m_{<e,h>,Q}$, by utilizing Eq.\ \eqref{eq1'} with $m_{<e,h>}$ data given for 4 ternary compounds \cite{ioffeGaAsSb, ioffeGaInAs, ioffeGaInSb, ioffeInAsSb}. Here $e$ is electron charge, $k_B$ is Boltzmann constant and $h$ is Planck's constant. Minority carrier capture cross section $\sigma=1.5\times10^{-19}{\text{ m}}^2$ and density of SRH traps $N_t=1.17\times10^{21} {\text{ m}}^{-3}$ are constant. Recombination coefficients $B_\text{rad}$ and $C_\text{Auger}$ are easily found for 4 binary compounds and Vegard's law $\{$Eq.\ \eqref{eq2'}$\}$ with zero bowing parameters is applied to obtain $B_{\text{rad},Q}$ and $C_{\text{Auger},Q}$. With every parameters known, carrier lifetime $\tau_{<e,h>,Q}$ are driven from Eq.\ \eqref{eq5'}.

Returning to optical property, we calculate dielectric function below the semiconductor's bandgap energy with Lorentz-Drude Oscillator model \cite{rakic1998optical}.
\begin{equation}\label{eq6'}
\epsilon(\omega)={\epsilon_{\infty}}\big(1+\frac{{\omega_{\text{LO}}}^2-{\omega_{\text{TO}}}^2}{{\omega_{\text{TO}}}^2-{\omega}^2-i{\omega}{\gamma}}-\frac{{\omega_{p}}^2}{\omega(\omega+i\Gamma)}\big)
\end{equation}

$\epsilon_{\infty}$, $\omega_{\text{LO}}$, $\omega_{\text{TO}}$, $\gamma$, $\Gamma=\frac{e}{m_{e}\mu_{e}}$, and $\omega_{p}=({\frac{{N_A}e^2}{{\epsilon_{0}}{\epsilon_{\infty}}m_{e}}})^{1/2}$ needs to be obtained for quaternary compound using Vegard's law. $m_{e}$ and $\mu_{e}$ are electron effective mass and mobility obtained from above and $\epsilon_{0}$ is permittivity of the vacuum. $\epsilon_{\infty,Q}$, $\omega_{\text{LO},Q}$, $\omega_{\text{TO},Q}$ and $\gamma_{Q}$ are driven from binary compound properties using Eq.\ \eqref{eq2'} with zero bowing parameters. $\Gamma_{Q}$ and $\omega_{Q}$ are calculated after that. With Eq.\ \eqref{eq3'} and Eq.\ \eqref{eq6'} together, it is able to obtain dielectric function of quaternary semiconductor for both above and below its bandgap energy. Furthermore, spectral optical property of a quaternary semiconductor with arbitrary composition $[x,y]$ was predicted utilizing Vegard's law $\{$Eq.\ \eqref{eq1'}, \eqref{eq2'}$\}$ to find model parameters and coefficients.

\newpage
\section*{3. Layer-by-layer absorption in tandem TPV cell}
The spectral radiative heat flux described in Fig.\ S3(a) shows the layer-by-layer absorption in the tandem TPV cell for the optimum case $\{$the configuration shown in Fig.\ 3(a) of the manuscript$\}$. The Au reflector and tunnel junction absorb marginal amounts of radiation in the above-bandgap spectral region, while we cannot ignore the absorption in the sub-bandgap spectral region. About half of the parasitic absorption loss $\{$see Fig.\ 5(c) of the manuscript$\}$ is occupied by the Au reflector and tunnel junction. On the other hand, as shown in Fig.\ S3(b), we can see the large absorption in the InAs substrate for the bare case, which is without ITO film and Au reflector from the optimum case. Even in the spectral region between $\omega_\text{g,bottom}$ and $\omega_\text{g,top}$, radiation that is not absorbed by the bottom cell is transmitted and absorbed into the substrate. Because of the large parasitic loss (50.4$\%$ of the total absorbed heat flux) mainly attributed to the InAs substrate absorption, the conversion efficiency becomes low compared to the optimum case. Since there is no Au reflector, multiple reflections do not occur inside the TPV cell, such that there is little absorption by the tunnel junction.

\begin{figure}[h]
\centering\includegraphics[width=17cm]{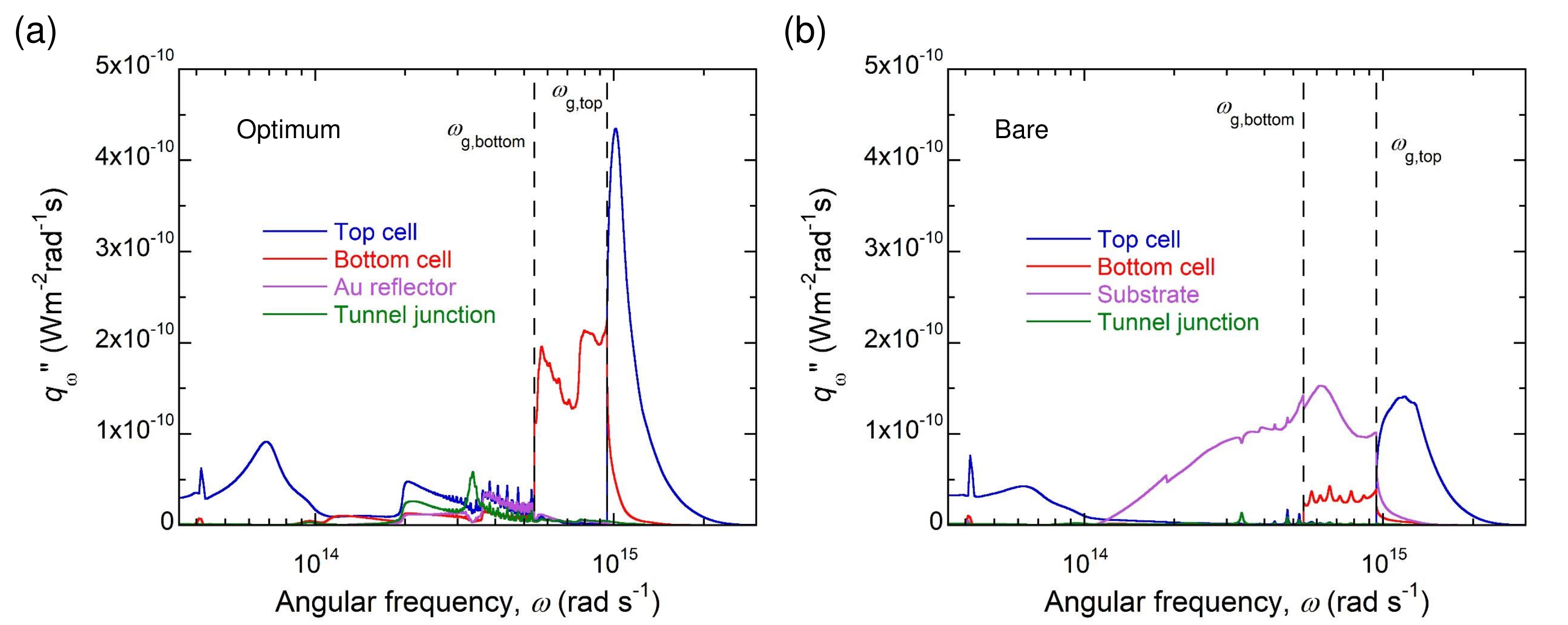}
\caption{\label{Figs2}Fig.\ S3: Layer-by-layer spectral radiation absorbed in the tandem TPV cell (a) for the optimum case and (b) for the bare case.}
\end{figure}

\newpage
\section*{4. Effects of ITO film and Au reflector}
The surface plasmon polariton (SPP) resonance due to the ITO-vacuum interface and the waveguide mode due to the Au reflector separately contribute to the main increment of the absorption in the top cell and bottom cell, respectively. In the manuscript, we deal with the optimum case with thin-ITO covered emitter and Au backside reflector, and the bare case without ITO film and Au reflector (see Figs.\ 3 and 4 of the manuscript). To specifically examine the individual effects of the ITO film and Au reflector on the radiative heat flux absorbed by the TPV cell, additional calculations are performed for the structure without the Au reflector or the ITO film from the optimum case. Figures S4 and S5 show the spectral radiative heat flux and the contour plot of $S_{\beta,\omega}$ including those cases. Figure S4(a) shows the radiative heat flux absorbed by the top cell for four cases. When there is no Au reflector, the spectral heat flux absorbed by the top cell is almost the same as the optimum case, whereas when the ITO film is removed, it becomes similar to the bare case. Comparing Fig.\ S5(h) and Fig.\ S5(k), we can physically see that the Au backside reflector has a marginal effect on the above-bandgap absorption of the top cell. Hence, we clearly demonstrate that the SPP resonance due to the ITO film mainly contributes to enhancing the top cell absorption for the optimum case. The radiative heat flux absorbed by the bottom cell calculated for the four cases is shown in Fig.\ S4(b). Because of the resonant emission near the bandgap frequency of the top cell due to the ITO film, the absorption into the bottom cell also increases in this spectral region. However, the employment of the Au reflector shows the dominant effect to increase the photocurrent generated in the bottom cell while minimizing thermalization loss by increasing the absorption close to the frequency corresponding to the bottom cell bandgap. As can be seen in Figs.\ S5(c),(f),(i), and (l), the waveguide mode can support the absorption enhancement to the bottom cell only when the electromagnetic field is confined by the backside reflector.

\begin{figure}[h]
\centering\includegraphics[width=17cm]{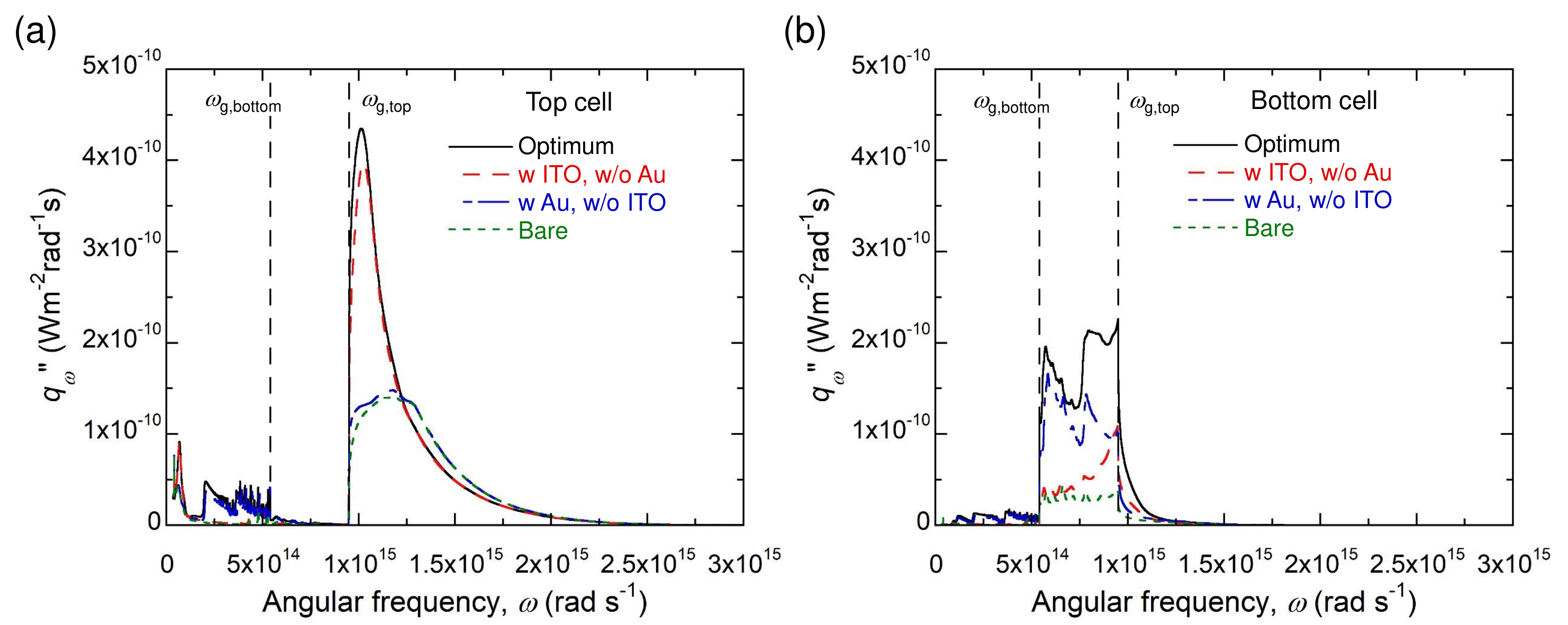}
\caption{\label{Figs2}Fig.\ S4: Spectral radiative heat flux for four cases: optimum, without Au reflector, without ITO film, and bare absorbed in (a) top cell and (b) bottom cell.}
\end{figure}

\newpage
\begin{figure}[h]
\centering\includegraphics[width=17cm]{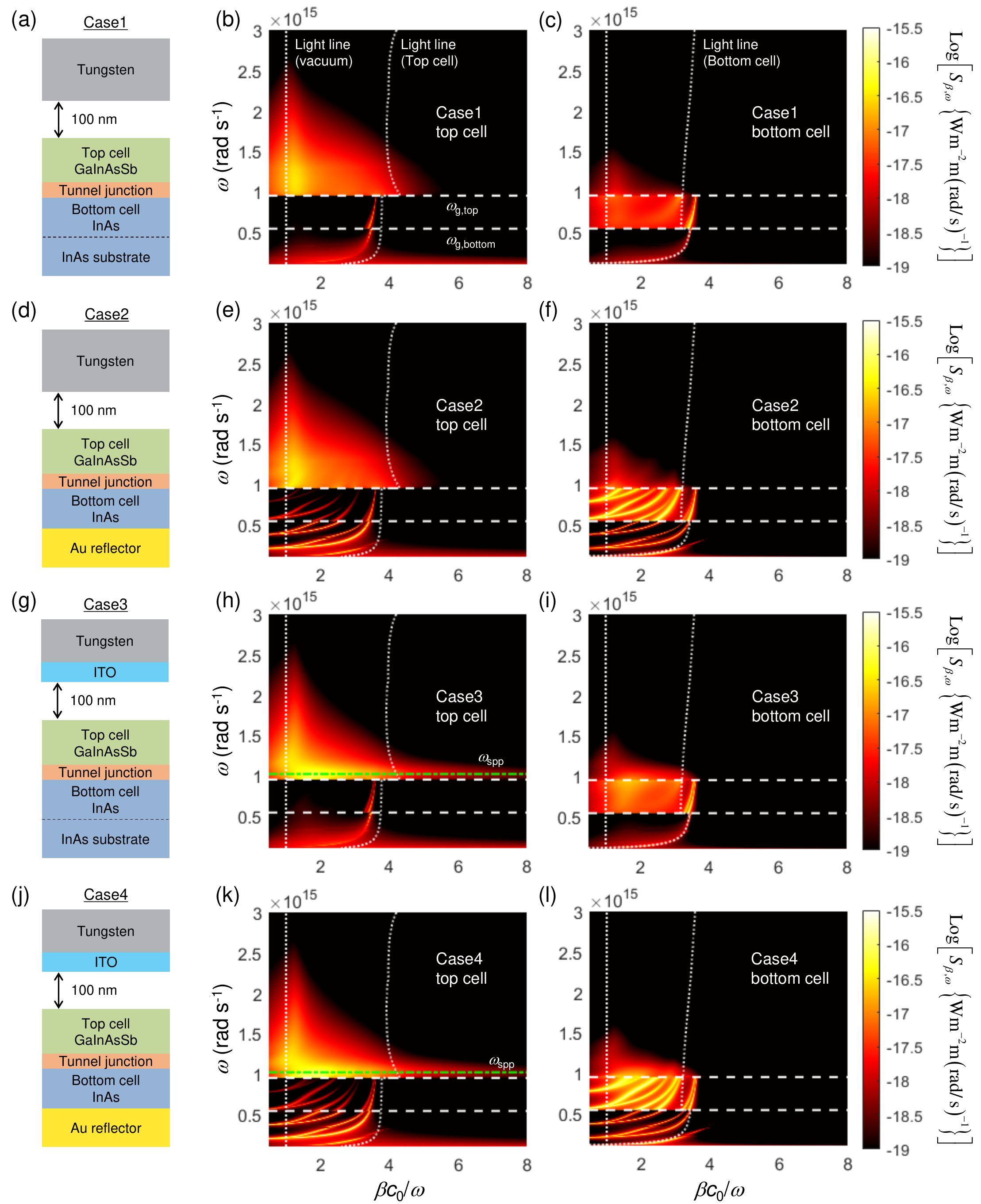}
\caption{\label{Figs2}Fig.\ S5: Contour plots of $S_{\beta,\omega}(\beta,\omega)$ absorbed in the top cell and the bottom cell for four cases: (a-c) bare case, (d-f) with Au reflector and without ITO film case, (g-i) with ITO film and without Au reflector case, and (j-l) optimum case.}
\end{figure}

\clearpage
\bibliographystyle{elsarticle-num-names}
\bibliography{Song_Bib.bib}